\documentclass[useAMS,usenatbib]{mn2e}
\usepackage[pdftex,pdfpagemode={UseOutlines},bookmarks,bookmarksopen,colorlinks,linkcolor={blue},citecolor={blue},urlcolor={cyan}]{hyperref}

\usepackage{txfonts}
\usepackage[T1]{fontenc}
\usepackage{graphicx}
\usepackage[lofdepth,lotdepth]{subfig}
\usepackage{array}

\usepackage{amsmath,amssymb}
\usepackage[english]{babel}
\usepackage{enumerate}
\usepackage{xspace}

\usepackage[usenames,dvipsnames]{color}

\usepackage{pythonhighlight}
\usepackage{booktabs}

\usepackage{xspace}
\usepackage{verbatim}
\usepackage{booktabs}
\usepackage{mathtools}
\usepackage{enumitem}
\usepackage{multicol,multirow}
\usepackage{aas_macros}

\DeclarePairedDelimiter\ceil{\lceil}{\rceil}
\DeclarePairedDelimiter\floor{\lfloor}{\rfloor}
\DeclarePairedDelimiter\abs{\lvert}{\rvert}
\usepackage{mathrsfs}
\providecommand{\spark}{{\tt Spark}\xspace}
\providecommand{\aspark}{\texttt{Apache} \texttt{Spark}\xspace}
\providecommand{\sql}{{\tt Spark SQL}\xspace}
\providecommand{\exact}{{\tt exact}\xspace}
\providecommand{\red}{{\tt reduced}\xspace}
\providecommand{\py}{{\tt python}\xspace}
\providecommand{\sca}{{\tt Scala}\xspace}
\providecommand{\sql}{{\tt Spark SQL}\xspace}
\providecommand{\hp}{{\tt HEALPix}\xspace}
\providecommand{\cs}{{\tt CubedSphere}\xspace}
\providecommand{\cobe}{{\tt COBE}\xspace}
\providecommand{\sa}{{\tt SARSPix}\xspace}
\providecommand{\angpix}{{\tt ang2pix}\xspace}
\providecommand{\pixang}{{\tt pix2ang}\xspace}
\providecommand{\neib}{{\tt neighbors}\xspace}
\providecommand{\LSST}{{\tt LSST}\xspace}
\newcommand{\Npix}{\ensuremath{N_\mathrm{pix}}\xspace}
\newcommand{\NpixD}{\ensuremath{N_\mathrm{pix}^D}\xspace}

\newcommand{\NpixJ}{\ensuremath{N_\mathrm{pix}^J}\xspace}

\newcommand{\nbase}{\ensuremath{\mathtt{nbase}}\xspace}
\newcommand{\nbaseD}{\ensuremath{\mathtt{nbase}_D}\xspace}
\newcommand{\nbaseJ}{\ensuremath{\mathtt{nbase}_J}\xspace}

\newcommand{\npairsX}{\ensuremath{N_\textrm{pairs}^X}}
\newcommand{\npairsR}{\ensuremath{N_\textrm{pairs}^R}}
\newcommand{\nside}{\ensuremath{\mathtt{nside}}\xspace}

\newcommand{\Rin}{\ensuremath{R_\mathrm{in}}\xspace}
\newcommand{\Rout}{\ensuremath{R_\mathrm{out}}\xspace}
\newcommand{\Rsqout}{\ensuremath{R_\mathrm{sq}^{out}}\xspace}
\newcommand{\Rsqin}{\ensuremath{R_\mathrm{sq}^{in}}\xspace}
\newcommand{\td}{\ensuremath{\theta_-}\xspace}
\newcommand{\tu}{\ensuremath{\theta_+}\xspace}

\newcommand{\imin}{\ensuremath{i_\mathrm{min}}}
\newcommand{\imax}{\ensuremath{i_\mathrm{max}}}
\providecommand{\sparkcorr}{{\tt SparkCorr}\xspace}

\providecommand{\colore}{{\tt CoLoRe}\xspace}
\newcommand{\citeg}[1]{\citep[e.g.,][]{#1}}
\def\q#1{\textquotedblleft{#1}\textquotedblright}
\newcommand{\SF}{\texttt{spark-fits}\xspace}
\newcommand{\safeurl}[1]{\href{#1}{#1}}
\newcommand{\bigO}[1]{\ensuremath{\mathcal{O}(#1)}}

\newcommand{\Ndata}{\ensuremath{\textrm{N}_\mathrm{data}}\xspace}

\providecommand{\text}[1]{\rm{#1}}

\newcommand{\begm}{\begin{pmatrix}}
\newcommand{\enm}{\end{pmatrix}}

\newcommand\ba{\begin{eqnarray}}
\newcommand\ea{\end{eqnarray}}
\newcommand\bea{\begin{eqnarray}}
\newcommand\eea{\end{eqnarray}}

\newcommand\be{\begin{equation}}
\newcommand\ee{\end{equation}}

\def\pmb#1{\setbox0=\hbox{#1}%
    \kern-.025em\copy0\kern-\wd0
    \kern.05em\copy0\kern-\wd0
    \kern-.025em\raise.0433em\box0}
\def\ltsima{$\; \buildrel < \over \sim \;$}
\def\gtsima{$\; \buildrel > \over \sim \;$}
\def\simlt{\lower.5ex\hbox{\ltsima}}
\def\simgt{\lower.5ex\hbox{\gtsima}}

\def\etal{{\it et al.}}

\def\p2Y{\;_2Y}
\def\m2Y{\;_{-2}Y}

\newcommand{\ev}{\ensuremath{\mathrm{\,e\kern -0.1em V}}\xspace}

\def\ie{{i.e.}\xspace}
\def\eg{{e.g.}\xspace}

\def\wrt{with respect to\xspace}

\def\to{\ensuremath{\rightarrow}\xspace}

\renewcommand{\refeq}[1]{Eq.~(\ref{eq:#1})\xspace}
\newcommand{\sect}[1]{Sect.~\ref{#1}\xspace}
\newcommand{\fig}{Fig.}
\newcommand{\change}[1]{{\bf #1}}
\newcommand{\JP}[1]{{#1}}
\usepackage{colortbl}
\definecolor{darkblue}{rgb}{0.0, 0.0, 0.55}
\usepackage{xcolor}
\usepackage[normalem]{ulem}
\title{Scaling pair count to next galaxy surveys}
\author[S. Plaszczynski \etal]{S. Plaszczynski, J.E. Campagne, J.
  Peloton, and C. Arnault \\
Universit\'e Paris-Saclay, CNRS/IN2P3, IJCLab, 91405 Orsay, France}

\date{\today}

\begin{document}

\maketitle

\begin{abstract}
Counting pairs of galaxies or stars according to their
distance is at the core of real-space correlation analyzes performed in astrophysics and cosmology.
Upcoming galaxy surveys (\LSST, \texttt{Euclid}) will measure properties of  billions of galaxies
challenging our ability to perform such counting in a minute-scale time relevant for the usage of simulations. 
The problem is only limited by efficient
access to the data, hence belongs to the big data category.
We use the popular \texttt{Apache Spark} framework to address it and design an
efficient high-throughput algorithm to deal with hundreds of millions to
billions of input data.
To optimize it, we revisit the question of nonhierarchical sphere pixelization
based on cube symmetries and develop a new one dubbed the "Similar
Radius Sphere Pixelization" (\texttt{SARSPix}) with very close to square pixels.
It provides the most adapted indexing over the sphere for all distance-related computations.
Using LSST-like fast simulations, we compute autocorrelation
functions on tomographic bins containing between a hundred million to
one billion data points. In each case we achieve the construction
of a standard pair-distance histogram in about 2 minutes, using a
simple algorithm that is shown to scale, over a moderate number of
nodes (16 to 64). 
This illustrates the potential of
this new techniques in the field of astronomy where data access is
becoming the main bottleneck. They can be easily adapted to other use-cases as
nearest-neighbors search, catalog cross-match or cluster finding.
The software is publicly available from https://github.com/astrolabsoftware/SparkCorr.
\end{abstract}


%%%%%%%%%%%%%%%%%%%%%%%%%%%%%%%%%%%%%%

%\tableofcontents

\section{Introduction}
Since the very beginning of large-scale structure studies
\citep{Peebles80}, counting the number of pairs of galaxies
according to their distance is the basis for the estimation of
two-point correlation functions in cosmology.
Including also today the measured ellipticities to infer the cosmic
shear provides invaluable information on the  dark energy at an epoch
it starts playing a sizable role on structure formation. 
The statistically most powerful way to analyze the data is through the
so-called \q{$3\times2$-points} method, which combines two galaxy
samples, a faint one with estimated shear distortions and a smaller
one with more stringent selection criteria \citeg{Kilbinger:2015}. 
One then performs both autocorrelations and cross-correlation between
the shears and the positions which allows not only to estimate
cosmological and nuisance parameters but also to carry out systematic tests.
\citeg{Mandelbaum:2018}.
There are in fact more than three correlation functions since the
galaxy population is split into redshift bins (the tomographic shells)
and correlations are performed between them which allows a model
independent data analysis still with negligible precision loss for
photometric surveys.

The latest tomographics surveys \citep{DES:2018,HSC:2019,KIDS:2020}
use shells with tens of 
millions of objects (for the faint sample). 
State-of-the-art algorithms, \texttt{treecorr} \citep{Jarvis:2004} or \texttt{CUTE} \citep{Alonso:2013}
are run to compute the shear-shear,
position-position and shear-position statistics. 
These algorithms are based on High Performance Computing (HPC) on
supercomputers and results are obtained within minutes on those
\bigO{10^7} datasets.

The next generation of galactic surveys,  the ground-based Vera C. Rubin Observatory
Legacy Survey of Space and Time (\LSST)\footnote{\safeurl{www.lsst.org}} \citep{LSST:2009} and
\texttt{Euclid} spatial
mission\footnote{\safeurl{www.cosmos.esa.int/web/euclid},
  \safeurl{www.euclid-ec.org}} \citep{Euclid:2011} will push the 
statistics much higher, measuring properties of billions of galaxies.
The planned tomographic shells will then contain hundreds of millions
of objects. To our knowledge no pair-counting algorithm can attain minutes
performances with such numbers.

Distance binned pair-counting is not a CPU intensive computation. It
is only limited by heavy combinatorics, \ie by efficient access to the
data.
It becomes then natural to inspect if \q{big data} methods can be of any help here.
A popular tool to analyze huge amounts of data is \aspark \citep{Zaharia:2012,Zaharia:2010} which is an efficient implementation of the original
\texttt{map-reduce} paradigms presented by \texttt{Google} \citeg{Dean:2008} 
and of several tools developed in the \texttt{Hadoop}
ecosystem \footnote{\safeurl{hadoop.apache.org}}.
We point out that this branch of computing, loosely known as
High Throughput Computing (HTC) relies on ideas that are very
different from the standard operative methods (e.g. in \texttt{fortran},
\texttt{C++}), using for instance the functional programming paradigm
that is getting popular these years due to excellent performances over
large datasets.

\spark is essentially used by private companies although some dedicated
systems using it were also successfully developed in astrophysics for instance to
organize data access \citep{AXS:2018,Astroide:2018}.
It is a multi-purpose engine that can deal not only with standard
multi-variable analysis with the \texttt{SQL}
module \citep{SparkSQL:2015}, but
also with large graphs (with some application to astrophysics in
\citep{Hong:2020}), machine learning \citep{Meng:2015} and streaming
capabilities (with a recent applications to an alert
broker\footnote{\safeurl{fink-broker.org}} in
\citet{FINK}).

Although not much used in science, the goal of this work is to show
that a tool like \spark can address some modern limitations due to
heavy data access and will become more and more pertinent in the next years.
Furthermore as will be shown, we try to convince users that beside
changing some habits, the learning curve is quite smooth and rapid.
With little work one can obtain the same performances
than dedicated software developed by experts for years.

In the following we will use only some very basic aspects of \spark, namely
\textit{dataframes} much popularized in astrophysics by the \py
\texttt{pandas} package\footnote{\safeurl{https://pandas.pydata.org}}.
A pedagogical introduction to the use of \spark dataframes in cosmology and
astrophysics is given in \citet{plaz:2019} (see also references
therein for more applications to astrophysics) but we will still
review some basics at the beginning of \sect{sec:method}.
Then, we will give the main ideas on how to reduce the pair-counting
combinatorics. They are not new, but we show how they can be adapted to \spark.

To improve on performances, we will the revisit the question
of sphere indexing in \sect{sec:pix} and propose an adapted new
pixelization scheme.
Although it was initially designed to
improve on our pair-counting algorithm, its scope is much more general. 
It can be used to pad the sphere with similar shape quadrangles and is
thus the best candidate for all distance-related computations. 
This section is essentially independent from the others.

Then we present in \sect{sec:results} the results obtained with a
logarithmic binning used by the {\tt DES} Collaboration, on  $10^8$ to
$10^9$ data points and detail performances before concluding on the
overall interest of the method. 
\ref{sec:app} gives some more information on some cubed-base pixelization
properties.

\section{Pair counting with \aspark}
\label{sec:method}

\subsection{Introduction to \sql}
\label{sec:sql}

\spark is a framework to work efficiently on a cluster of machines
not necessarily designed with (expensive) high speed connections, what is loosely called a data-center. It is
based on a \JP{driver-executor} architecture:  one machine (the driver) sends
instructions to a set of workers (executors), that perform some tasks and send back
some (reduced) results that are then combined by the driver. One
could say that the big data approach is about "sending computation to
the data" while HPC does it the reverse way, move the data to the compute units.
The essential feature here is then how data are stored and also
preserved through multiple copies, on each worker. This is called data
\textit{partionning}:
the goal is to achieve some locality to avoid
inter-nodes communications and it depends on the problem being studied.
Here for instance we will partition our dataset according to some pixel
number meaning gathering points (galaxies) within the same region of
the sky on the same workers.
\JP{Beyond} partitioning, an essential \JP{feature of \spark is the ability to} put the data in 
the executors' Random Access Memory \JP{hereafter, the
  \textit{cache}}. This \JP{boosts} performances considerably and for
the user it is equivalent to having a \JP{single} ``
machine'' with hundreds of Gigabytes of memory.
While \JP{the implementation under the hood can be quite} complex, on the user side it is not more complicated
than calling the \texttt{partition()} or \texttt{cache()} \JP{dataframe methods}.

\spark is a full framework with many functionalities, but we will be
concerned here only with dataframes which are part of the so-called
\sql module. Dataframes are not specific to \spark nor even coming
from the big-data field (one can find
similar ideas since the 70's in High Energy Physics where they are
called "n-tuples"). What \spark does is implement efficiently some standard
operations on them \JP{in a distributed environment}.

A \textit{dataframe} can be thought of simply as a table of values (a dataset of rows),
with some named columns. In our case the columns are
2 angles on the sky ("RA" and ``DEC") and some identifier ("id", a long
integer).

It is read from a file either in \texttt{FITS} format (using the \SF
connector, \citet{SparkFITS:2018}), either natively as plain text or
in \texttt{parquet} format, the latter being the choice we made. We will call
this dataframe the \textit{source}. Let us now review some important
operations with dataframes that we will use.

The most obvious one is to \texttt{filter} the dataframe meaning
applying conditions depending on some column values (as applying
cuts) to obtain a reduced one.

Another important operation is the \texttt{groupBy}. For a
given dataframe one can group rows according to some criteria and perform
some further operation of the grouped array to end up with some value
(as the number of its elements). To illustrate it let us see how
to obtain a histogram for some column as for instance the ``RA'' angle.
First one adds a new column with the bin number corresponding to each RA value.
Then one \texttt{groupBy} this bin number and count the number of
elements to estimate the frequency within each bin.
Here is how it would look like with \spark in \py: 

\begin{python}
# read the dataframe from a parquet file
# NB: files are available online
# see "data availability" section
src=spark.read.parquet("tomo100M.parquet")  
# what's in it?
src.show(3)
+---------+----------+
|       RA|       DEC|
+---------+----------+
|180.77408|-40.784428|
|180.78752| -40.85317|
|180.75848|-40.799908|
+---------+----------+
# let us do a histogram on "RA" (10 degrees bins)
# 1. add the bin number
h=src.withColumn("bin",(df.RA/10).astype('int'))
# check
h.show(3)
+---------+----------+---+
|       RA|       DEC|bin|
+---------+----------+---+
|180.77408|-40.784428| 18|
|180.78752| -40.85317| 18|
|180.75848|-40.799908| 18|
+---------+----------+---+
# 2. bin counting: groupby bin number 
# and count the number of elements in each group
h=h.groupBy("bin").count()
# check
h.show(3)
+---+-----+
|bin|count|
+---+-----+
| 26|61288|
| 12|60965|
| 22|89625|
+---+-----+
# then we may sort according to the bin number, 
# add bin centers etc.
\end{python}

With two dataframes, one can perform a \textit{join} operation: this
means creating a dataframe where each row is obtained by ``gluing''
together two rows from the previous ones, according to a common
column value. For
instance let us add to each row of the source some pixel number. If we
perform a join operation between the source \textit{and itself} based on the
pixel number, we will end up with a dataframe containing all pairs
having the same pixel number. 
We always use some \textit{inner-} join operations which means
that the two rows must exist in each dataframe to perform the gluing.
To remove auto-pairs and reverse-id pairs we filter the output
requiring that the first id be lower than the second. 
An example of such operation is show in Table \ref{tab:pairs}.

\begin{figure}
  \centering
  \includegraphics[width=\linewidth]{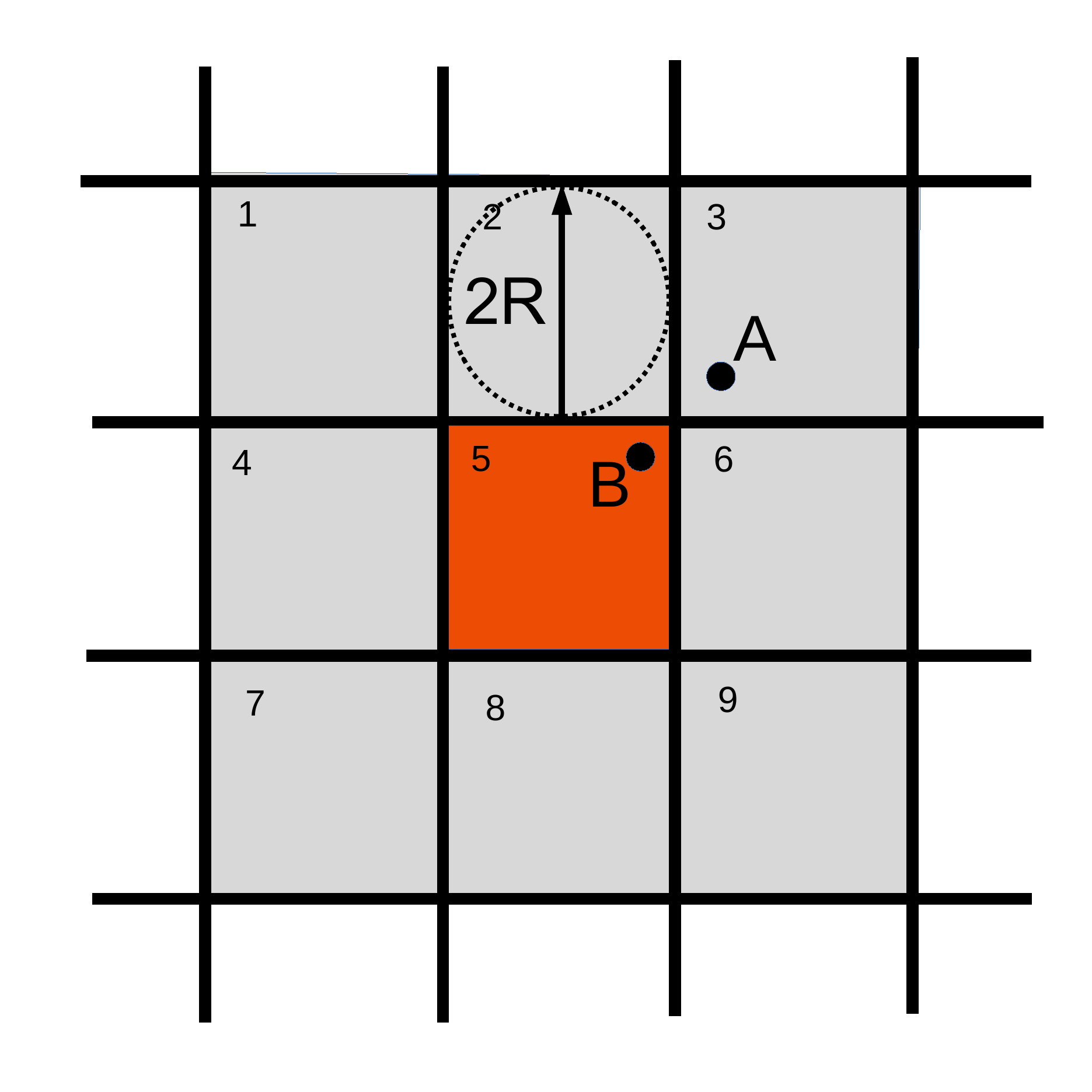}
  \caption{  \label{fig:grille}
    Illustration of pair building on a pixel grid (square
    lattice here). The grey pixels corresponds to the neighbours of
    the red one and will be denoted \textit{duplicates} (including the
      red one itself). The numbers denote some pixel indices.
      The radius involved for the choice of the pixel size
      is shown and corresponds to the inner radius of the pixel.}
\end{figure}

\begin{table}
\centering
\begin{tabular}{|l|l|l|l|}
\hline
  RA &    DEC &  id &  ipix \\
\hline
180.77 & -40.78 &   0 &   0 \\
180.79 & -40.85 &   1 &          0 \\
180.76 & -40.80 &   2 &          0 \\
180.65 & -41.00 &   3 &          1 \\
180.55 & -41.01 &   4 &          1 \\
180.46 & -40.81 &   5 &          1 \\
\hline
\end{tabular}

\hskip.3cm
\begin{tabular}{|l|l|l|l|}
\hline
   RA2 &   DEC2 &  id2 &  ipix \\
\hline
180.77 & -40.78 &    0 &          0 \\
180.79 & -40.85 &    1 &          0 \\
180.76 & -40.80 &    2 &          0 \\
180.65 & -41.00 &    3 &          1 \\
180.55 & -41.01 &    4 &          1 \\
180.46 & -40.81 &    5 &          1 \\
\hline
\end{tabular}

\vskip.4cm
inner-join\\
$\big\Downarrow$
\vskip.5cm
\begin{tabular}{|l|l|l|l|l||l|l|}
\hline
ipix &     RA &    DEC &  id &    RA2 &   DEC2 &  id2 \\
\hline
         0 & 180.77 & -40.78 &   0 & 180.79 & -40.85 &    1 \\
         0 & 180.77 & -40.78 &   0 & 180.76 & -40.80 &    2 \\
         0 & 180.79 & -40.85 &   1 & 180.76 & -40.80 &    2 \\
         1 & 180.65 & -41.00 &   3 & 180.55 & -41.01 &    4 \\
         1 & 180.65 & -41.00 &   3 & 180.46 & -40.81 &    5 \\
         1 & 180.55 & -41.01 &   4 & 180.46 & -40.81 &    5 \\
\hline
\end{tabular}

\caption{\label{tab:pairs}
Illustration of the inner-join operation. The upper left
  table represent the source dataframe with some angular positions
  (RA/DEC), an identifier (id) and a pixel number
  ("ipix"). The upper right is just a copy of it
  with the columns renamed but for "ipix". 
The lower table is the result of applying
an inner-join operation based on "ipix" and filtering the output
according to $id<id2$. It contains all the pairs of points with the
same pixel value.}
\end{table}
This is a very optimized way of building all
pairs of nearby points but that neglects the case where
each point lies across a pixel boundary (as A and B on Fig.\ref{fig:grille}).
To resolve it we will use a trick due to \citet{BrahemYZ18}. Starting
from the \textit{source} dataframe we do not simply copy it as in
Tab.\ref{tab:pairs}. Instead each row is replicated and indexed not only by its pixel
number but also by its 8 neighboring pixel numbers. We will call this
9 times larger dataframe, the \textit{duplicates} one. 
Consider points A and B on Fig.\ref{fig:grille} with their pixel
number.
In a symbolic form we can write the \textit{source} dataframe as :
$(A,3),(B,5)$. In the \textit{duplicates} dataframe the B point is repeated 9
times : $(B,1),(B,2),(B,2),(B,3),(B,4),(B,5),(B,6),(B,7),(B,9)$.
Consider what happens in an inner-join between the source and its
duplicates: only the $(A,3),(B,3)$ pair has a common index and will be
retained in the join operation. In this way
we build the pairs between all points lying in the same pixel but also in the
neighboring ones. Of course many pairs will have a too large distance now.
But once the pair dataframe is available one can compute exactly the
distance between them and filter precisely on it.

\subsection{Binning}
\label{sec:binning}
Counting pairs according to the distance between objects in order to
compute correlation functions is always performed within a
histogram defined by a given \textit{binning}.
As an illustration, we will use later the one used by the
DES Collaboration in their first year $3\times2$ point analysis \citep{DES:2018}
which consists of 20 bins uniformly spaced on a logarithmic scale from 2.5\arcmin to
250\arcmin. Results will be discussed in detail in
\sect{sec:results} and we only emphasize here that because of the
logarithmic scale the bin width \textit{increases} with the bin number.
We will focus on
the autocorrelation of the shear sample which has the most intensive
combinatorics. Since computing shear-related quantities is not
CPU-limiting we just study hereafter the performance of pair counting.

\subsection{Building pairs}
\label{sec:exact}
Working on more than \bigO{10^8} data points, the raw number of pairs
scaling as $N^2$ precludes any hope for computing all the pair-distances.
Fortunately, there exists some classical methods to reduce combinatorics and
we show how they can be implemented in \spark.

The fist obvious one is to compute distances only for pairs that are
below the upper bin cutoff.
To this end we add some pixel index with a suitable size that will be discussed later.
Then, as explained in \sect{sec:sql} we use an \texttt{inner-join}
operation betwen the \textit{source} and the \textit{duplicates}
dataframes on it.

Let us consider that there are \NpixJ pixels over the sphere.
Then assuming each pixel holds $\Ndata/\NpixJ$ values in average,
the number of pairs \footnote{We use an \q{X} index
  since it uses an exact distance computation.} reduces to
\begin{align}
  \npairsX\approx& \NpixJ\times\dfrac{1}{2}\left(\dfrac{\Ndata}{\NpixJ}\right)^2 =\dfrac{\Ndata^2}{2\NpixJ}.
\end{align}

However since we will be considering pixelization with essentially 8
neighbours, the \textit{duplicates} dataframe is 9 time larger 
than the
input data. Then the combinatorics increases to:
\begin{align}
\label{eq:npairsX}
  \npairsX\approx\dfrac{9^2\Ndata^2}{2\NpixJ}.
\end{align}

We see that we want the largest possible number of pixels to decrease
combinatorics, i.e. the most compact padding of the sphere. This will be the topic of
\sect{sec:pix}.
We will call this pixelization the
\textit{joining} one and since this method calculates exact distances, we refer to it hereafter as the "exact method".

\subsection{Reducing the data}
\label{sec:reduce}
\begin{figure}
  \centering
    \includegraphics[width=\linewidth]{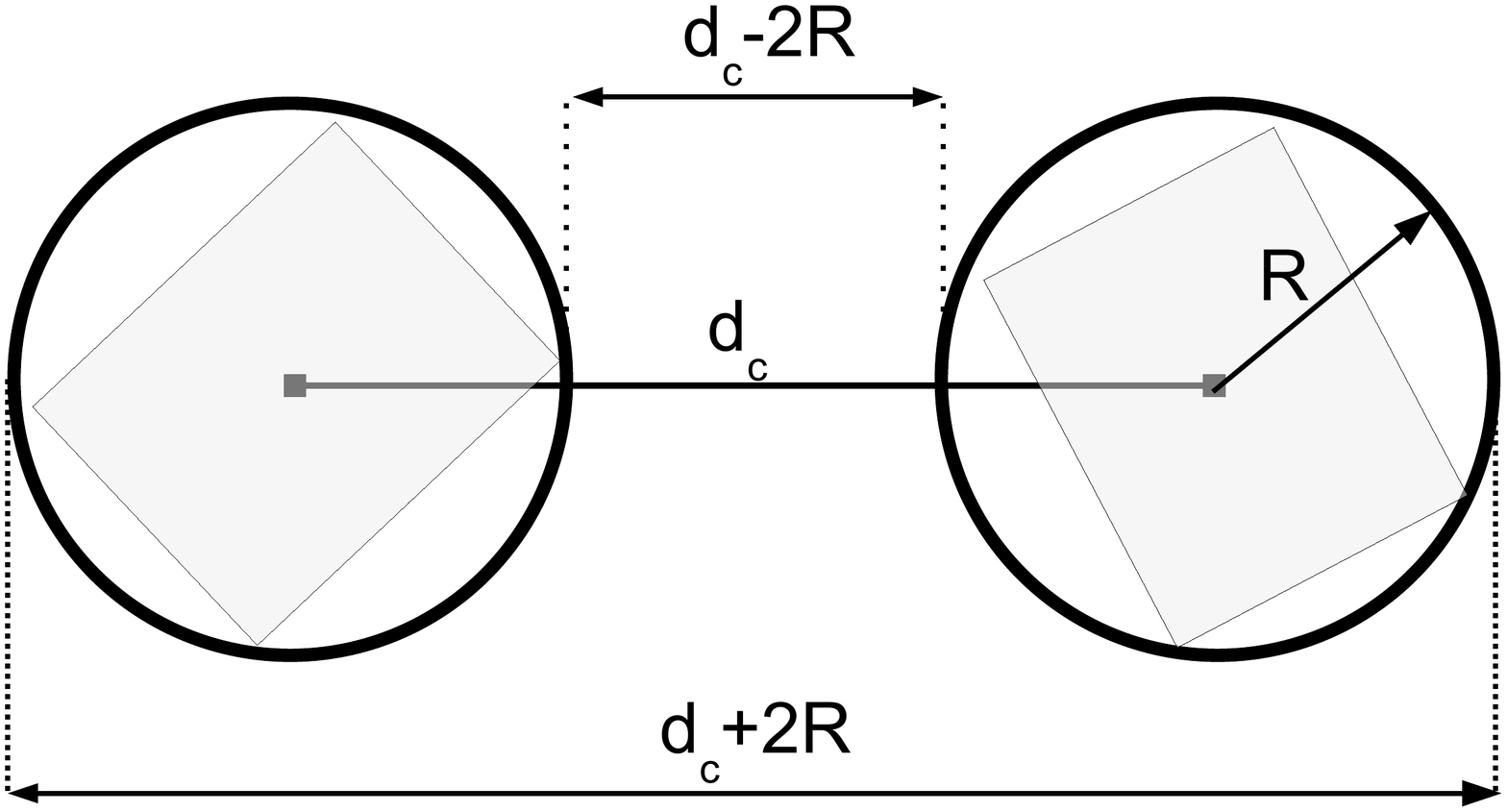}
    \caption{\label{fig:boule}
      Sketch to illustrate the pixel size involved in data compression. Pixels (gray quadrangles) are
    contained in their \textit{circumscribed} circle and their centers
    are distant by $d_c$. The angular distance between any pair of points
    in each pixel is
    contained within $[d_c-2R,d_c+2R]$. The radii involved here are
    the pixel outer ones.}
\end{figure}

Suppose we can group the data into areas on the sky
where all the points are below some distance $R$ from their center.
A first one has $N_1$ elements and the second one
$N_2$. Calling $d_c$ the distance between the two area, the
separation between any pair of points can be written as (see \fig\ref{fig:boule}) 
\begin{align}
  r_{ij}=d_c+\epsilon
\end{align}
where $\epsilon$ is a random variable satisfying
\begin{align}
  \abs*{\epsilon}<2R.
\end{align}

Let first consider a linear binning starting at $r_0$ of width $w$.
The bin number is obtained from 
\begin{align}
\label{eq:binning}
  \floor*{\dfrac{r_{ij}-r_0}{w}}\simeq \floor*{\dfrac{d_c-r_0}{w}}+\floor*{\dfrac{\epsilon}{w}}
\end{align}
$\floor{}$ being the floor operator. This is not a strict equality
since $\epsilon$ introduces some smearing along the bin borders, but
is a reasonable approximation if $\epsilon \ll d_c-r_0$ and will be
tested later (\sect{sec:crosschecks}).
Then if $R$ satisfies
\begin{align}
\label{eq:bw}
  2R<w
\end{align}
the last term of \refeq{binning} is zero. 
Then, up to a tiny effect across the boundaries, 
all pairs have their angular separation falling withing the same bin,
so their contributions can be immediately computed as $N_1 N_2$.

Concerning the logarithmic binning scheme, as the bin width increases
with the angular separation, if we choose for $w$ its very first
value, \refeq{bw} is automatically satisfied for all the others.

To ensure that all point angular distances lie below a given radius, we use 
a new pixelization with a pixel size adapted to \refeq{bw}.
We will refer to it in the following as the \textit{reducing} pixelization.

We replace our input data by a new dataframe consisting of pixels, 
using their center as position and keeping their number count. This is
performed efficiently with a \texttt{groupBy} operation (see \sect{sec:sql}).
We then proceed following \sect{sec:exact}
changing the binning part by the
cumulative count of the product numbers.

When it can be applied, \ie when first the bin width is large enough that the
number of pixels \NpixD is lower than the input data size, it is a
very efficient compression scheme since the number of pairs decreases
quadratically and \refeq{npairsX} in this case becomes  

\begin{align}
\label{eq:npairsR}
  \npairsR\approx\dfrac{9^2(\NpixD)^2}{2\NpixJ}.
\end{align}

We will call this method the \red one.

\subsection{Algorithm overview}
\label{sec:overview}
We have now all in hands to design the pair counting algorithm with
\spark. We consider some given binning and first work out the first
bins with the \exact algorithm. The discussion on the number of
bins effectively treated is deferred to a concrete example in
\sect{sec:results}. We divide the algorithm into three main parts.
%\begin{description}[align=left]
\begin{enumerate}[noitemsep]
\item \textit{Source:}
\begin{itemize}[noitemsep]
%\begin{description}[align=left]
\item [S1]: read the input data angular coordinates into the
  \textit{source} dataframe, convert to Cartesian ones and add an
  increasing identifier on al the data.
\item [S2]: add a \texttt{joining} pixel index.
\item [S3]: do a repartition of the dataframe according to the pixel index and
  push it to the workers cache.
\end{itemize}
%\end{description}

\item \textit{Duplicates:}
\begin{itemize}[noitemsep]
\item [D1]: construct a new dataframe from the \textit{source} 
by re-indexing it to all the neighboring pixels.
\item [D2]: concatenate \textit{source}.
\item [D3]: do a repartition of the dataframe according to the pixel
  index and push it to the workers cache.
\end{itemize}

\item \textit{Join:}
\begin{itemize}[noitemsep]
\item [J1]: perform the (inner) \texttt{join} between the source and its
  duplicates to create all pairs.
\item [J2]: \texttt{filter} pairs for which the first identifier is
  lower than the second one, compute the distances and only keep
  distances that are below the upper bin cutoff.
\item [J3]: assign bin number to each distance and count.
\end{itemize}
\end{enumerate}
%\end{description}

The \textit{source} part is mainly limited by reading the data
from disks (I/O), the 
\textit{duplicates} part is more concerned by the network finite
bandwidth and the \textit{join} part by
accessing the data and doing actual computations.

The \red method is applied for the the following bins.
It is essentially the same than the previous one but adds a preliminar
stage to transform the input data:
\begin{enumerate}[noitemsep]
\setcounter{enumi}{-1}
\item \textit{Reduction:}
\begin{itemize}[noitemsep]
\item [R1]: read and add a new index to the data (reducing pixelization).
\item [R2]: group the points according to the R1 index and keep the
  number count (\texttt{groupBy}+\texttt{count}).
\item [R3]: compute and store the centers of the pixels in a new dataframe.
\end{itemize}
\end{enumerate}

Then we proceed according to the previous 1-3 steps using this dataframe
as our data, with a tiny modification to \textit{J3} since the bin numbers are
now computed from the product of the number of galaxies.

Before illustrating how it performs on a concrete case (\sect{sec:results})
we need to address the question of an adapted spatial space
indexing, \ie revisit the ancient question of sphere pixelization.

\section{Best spatial sphere indexing}
\label{sec:pix}

The indexing of the unit sphere enters in two different places in our method.

\begin{enumerate}
\item  Joining index (\textit{S2}) : the indexing scheme allows
  computing angular distances only
  for pairs that are below some upper \tu cut. 
  One needs a pixel size large enough that any pair
  of points within it is below the largest angular size of the
  problem. We aim to get the total number of pixels \NpixD to be as
  large as possible (\refeq{npixj}).
\item Data reduction index (\textit{R1}): in this step we replace our
  data by pixels when 
  their size is smaller than the smallest bin width of the problem.
  According to \refeq{npairsR} the goal is to choose the total number
  of pixels \NpixD as low as possible.
\end{enumerate}

As shown on Fig.\ref{fig:grille}, the \textit{S2} step concerns the
pixel \textit{inner radius}, defined by the largest \textit{inscribed}
circle, while the \textit{R2} step concerns its \textit{outer radius}
(\fig\ref{fig:boule}) 
defined by the smallest \textit{circumscribing} circle.
Depending on the tiling, pixels have different shapes so that
the inner and outer radii vary accordingly. 
For a given angular distance scale one must then use 
the minimal (inner case) or maximal (outer case) values in order 
to be valid everywhere.

Our factor of merit is then to have a high \textit{padding} of the sphere
leading to a high (resp. low) value of \NpixJ (resp. \NpixD).
In other words, we search for a pixelization scheme where the inner and outer
pixel radii vary little over the sphere.
We highlight in the following sections the search for this ideal indexing
focusing only on the inner and outer radius variations and 
defer some other results to \ref{sec:app}.

\subsection{\hp}
\label{sec:hp}

We first examine the  
\hp pixelization scheme \footnote{\safeurl{https://healpix.jpl.nasa.gov/}}
\citep{2005ApJ...622..759G} which is popular in
astrophysics and cosmology and provides highly optimized libraries
including the \texttt{java} language that can be natively interfaced
to \spark\footnote{\safeurl{https://github.com/cds-astro/cds-healpix-java}}.

\begin{figure}
  \centering
   \includegraphics[width=\linewidth]{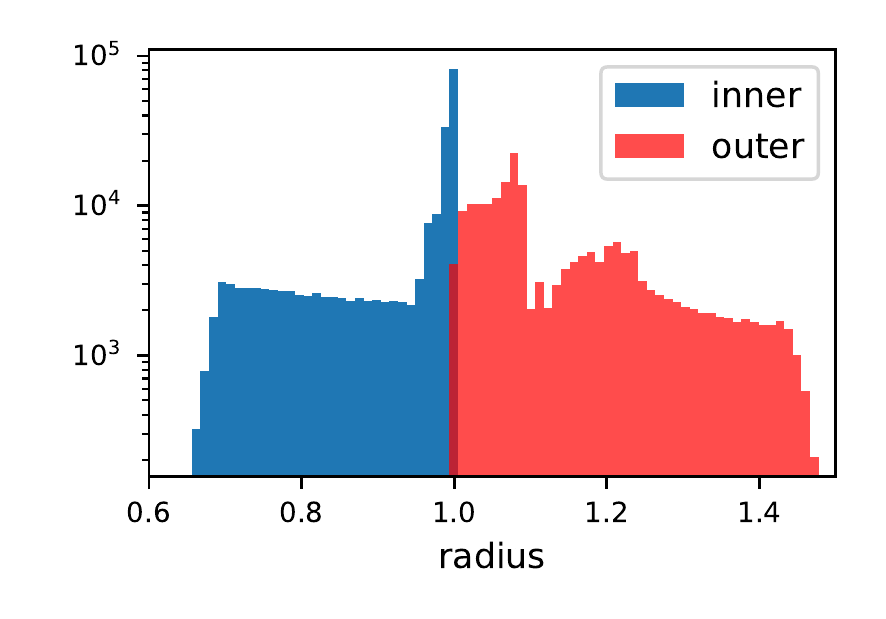}
  \caption{Histogram of the inner (in blue) and outer (in red)
    radii for \hp pixels in units corresponding to exact squares (\ie
    would be one for squares).
    We use a logarithmic scale to emphasize the tails and determine
    the lower (\Rin) and upper (\Rout) values.}
    \label{fig:hp}
\end{figure}

However \hp is designed for the efficient computation of
power spectra on the sphere which is not our concern here.
It has strictly equal-area pixels that are distributed on
constant latitude lines in order to perform fast spherical harmonic
transforms. As a result, pixels have quite a varying shape over the
sphere as will be shown next.
We study the pixel shapes from their boundaries provided by a \hp
function and compute the inner and outer radius distributions in the
following way. For each pixel, we determine the cartesian distance 
between the center and its 4 corners and take the maximal value as the
outer radius, while for the inner radius, we take the minimal value of
the length of the heights on each side.
We use a \nside=128 resolution parameter but make our
results independent of it by normalizing distances to those of exact
squares covering the full sky that would then have an inner and outer
radius of 
\begin{align}
\label{eq:Rsq}
  \Rsqin=\sqrt{\dfrac{\pi}{\Npix}},\quad
  \Rsqout= \sqrt{2} \Rsqin
\end{align}
In this way, deviation from unity shows how much the pixel is
square-like. \fig\ref{fig:hp} shows the results.
Both distributions deviate substantially from unity showing that the pixel
shape is far from being square.
For our indexing problem, what
matters is the minimal value of the inner radius (\Rin) and the
maximal outer radius one (\Rout) which is

\begin{align}
\label{eq:hp}
  \dfrac{\Rin}{\Rsqin}=0.65,\quad \dfrac{\Rout}{\Rsqout}=1.46.
\end{align}

\subsection{Homogeneous indexing}
\label{sec:needs}

What we are interested in is indexing the sky for an efficient access of 
neighboring points according to some angular distance. 
If the shape of pixels varies
too much, we are forced to use pixels that are in average too large  
to ensure we do not miss some neighbors on a
part of the sky and the padding is unsatisfactory.

This is worsened by using hierarchical decomposition. For instance 
\hp is based on the recursive subdivision of 12 equal-area base pixels, so that
the number of pixels follows $\Npix= 12~\nside^2$ where \nside must be
\textit{some integer power of 2}.
Then for a given angular distance cut, not only one has to use too large
pixels, but their size is, furthermore, artificially increased by the
fact that the resolution must be some power of 2 , which yields a
rapid growth of the number of pixels.
Therefore pixelizations as \hp, which were designed to perform power
spectra estimation, are not well adapted to our indexing needs.

We then look for a more homogeneous sphere indexing with the following most
constraining properties:

\begin{enumerate}
\item nonhierarchical decomposition,
\item small pixel inner/outer radius variations,
\item efficient \q{\angpix} possibility, \ie fast access to pixel index
  given a direction on the sky,
\item small (and about constant) number of pixel neighbors with
  efficient access to their indices.
\end{enumerate}
It is worth to mention however that we do not require a strict
equal-area pixelization.

As a matter of fact, nonhierarchical sphere tilings are rare nowadays. This
excludes some modern pixelization schemes used in geoscience as the \texttt{S2}
\footnote{\safeurl{https://s2geometry.io}} or
\texttt{H3}\footnote{\safeurl{https://h3geo.org/docs}} libraries (point 1).

There exists a vast literature on the subject of locating 
nodes on a sphere according to some quality criteria (for a review see
\eg \citet{hardin2016}).However pixels are most often constructed from
the nodes Voronoi mesh which precludes an efficient \angpix
implementation in \spark (point 3).

A priori the most interesting pixelization would be based on hexagons,
as the famous soccer ball scheme, since hexagonal pixels
are very close to circles and would then show a very small radius
variation. Unfortunately, it can be shown from the Euler
characteristic that it is impossible to pave entirely the sphere with only
hexagons; they must be complemented by exactly
12 pentagons which have unfortunately a much smaller area 
(0.66 of the hexagon ones).
For a given angular distance cut, we would then need to adapt our tiling
to the size of these pentagons losing the benefit of using
hexagons (point 2). 

We also exclude zonal equal array pixelizations, as the \texttt{IGLOO}
one \citep{IGLOO}, since the poles are contained within a single pixel
with a large number of neighbors (the number of meridians).
They would then add artificially a large number of neighbors to the
total pool which fails our 4th point.

Then we consider platonic solid projections on the sphere
which turn to be well adapted to the pair counting problem.
We revisit one of the simplest method based on the projection of 
points lying on the faces of a sphere inscribed cube, sometimes dubbed 
as "cubed sphere".

\subsection{Cubed-Spheres}

\begin{figure}
  \centering
  \subfloat[]{
    \includegraphics[width=.5\linewidth]{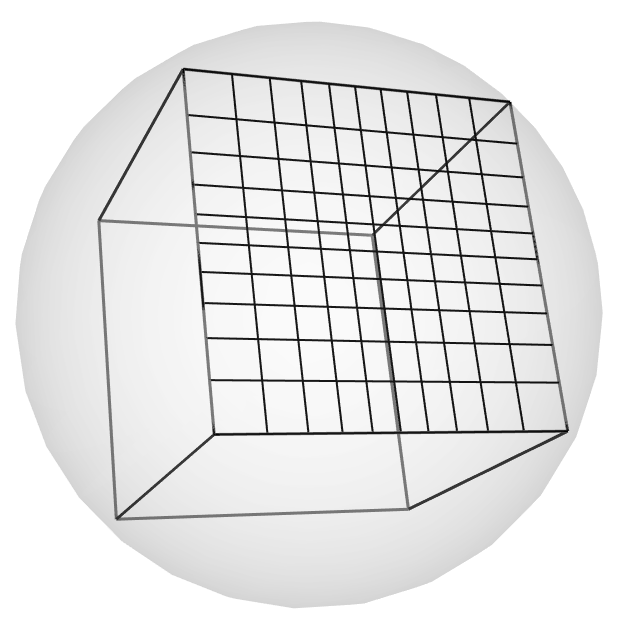}
    \label{sub:a}
  }
  \subfloat[]{
    \includegraphics[width=.5\linewidth]{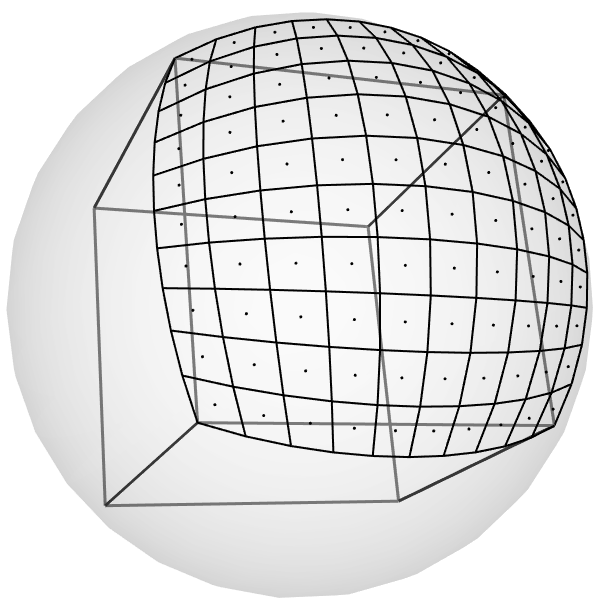}
    \label{sub:b}
  }\\
  \subfloat[]{
    \includegraphics[width=.5\linewidth]{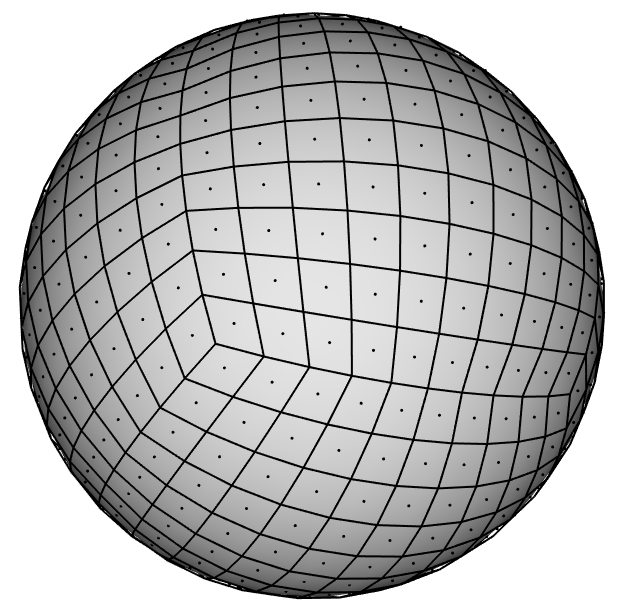}
    \label{sub:c}
  }
  \caption{Steps in the construction of a cubed sphere.
    (a) A cube is inscribed within a unit sphere, and a grid of nodes
    is created on the face (here with equal angles to the center). (b) Nodes are
    radially projected onto the unit sphere. Four connected nodes define a
    pixel their barycenter defines the pixel center. (c)
    The process is repeated on each face using cube symmetries. }
\label{fig:cubed}
\end{figure}

The construction of pixels on the sphere by radially projecting a grid of
points from the face of a cube (see \fig\ref{fig:cubed}) is a
classical method for instance well described in \citet{galerkin2005}.
The points can be positioned on a regular grid but a better strategy, still keeping
simplicity, is to place them at equal angles from the origin
\citep{Rancic}. This is our first discussed scheme. Next, a more
sophisticated procedure will be revisited and evaluated. 

\subsubsection{The equiangular case}
\label{sec:cs}

The nodes shown on \fig\ref{sub:a} are located in the local plane at position
\begin{align}
(x_i,y_j)=\tfrac{1}{\sqrt{3}}(\tan \alpha_i,\tan \alpha_j)  
\end{align}
where the $\alpha_{i,j}$ values represent the angle from the center
and are taken from a regular grid of \nbase points over $[-\pi/4,\pi/4]$.

The number of pixels per face is thus $\nbase^2$ and their total
number is
\begin{align}
\label{eq:Npix}
\Npix=6~\nbase^2.
\end{align}

The decomposition being nonhierarchical this number can be finely
tuned since \nbase can now be \textit{any integer number}.

\begin{figure}
  \centering
   \includegraphics[width=\linewidth]{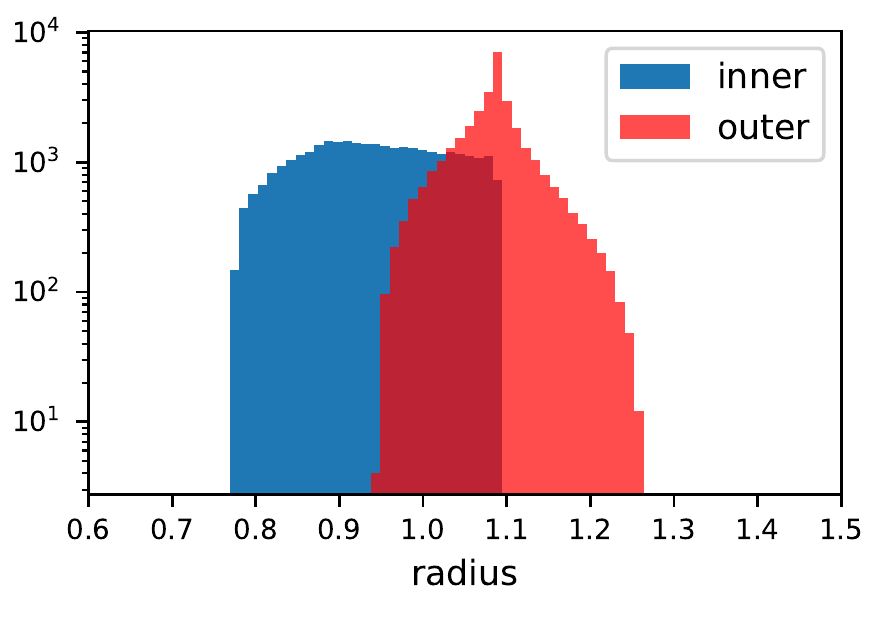}
  \caption{Histogram of the inner (in blue) and outer (in red) pixel
    radii for the \cs pixelization
    normalized to values for a square.}
    \label{fig:cs}
\end{figure}

The distributions of the inner and outer radius for this
pixelization is shown on \fig\ref{fig:cs}.
Compared to \fig\ref{fig:hp}, both distributions have clearly shrunk, the
shape of the pixels is closer to being square. We obtain
\begin{align}
\label{eq:cs}
  \dfrac{\Rin}{\Rsqin}=0.77,\quad \dfrac{\Rout}{\Rsqout}=1.26 
\end{align}

We have checked that this result is independent of \nbase.

\subsubsection{The \cobe quad-cube}
\label{sec:cobe}

The \cobe cosmological microwave background pioneer mission
developed an interesting sphere tiling known as the
"quad
cube"\footnote{\safeurl{https://lambda.gsfc.nasa.gov/product/cobe/skymap\_info\_new.cfm}}
which is a \cs projection but with a more elaborate placement of
points on the face of the cube. Based on an early work
by \citet{EPRF1975}, it proceeds by mapping a regular grid of
points with two invertible polynomials build
to ensure almost equal-area pixels after the projection.

\begin{figure}
  \centering
   \includegraphics[width=\linewidth]{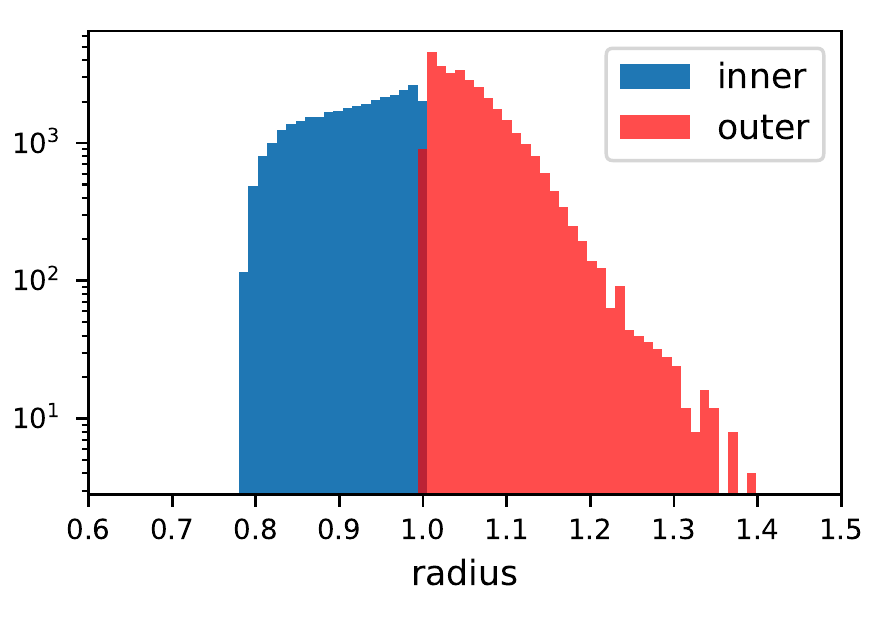}
  \caption{Histogram of the inner (in blue) and outer (in red) pixel
    radii for the \cobe pixelization
    normalized to values for a square.}
    \label{fig:co}
\end{figure}

We have thus implemented these polynomials in our grid construction
and show the effect on the pixel radii on \fig\ref{fig:co}. We obtain
\begin{align}
\label{eq:co}
  \dfrac{\Rin}{\Rsqin}=0.78,\quad \dfrac{\Rout}{\Rsqout}=1.39
\end{align}

The outer radius distribution now shows
an asymptotic tail that we traced to be related to projected pixels
near the corners of the cube that have both a larger surface area and
greater ellipticity (see \ref{app:cobe} for details). 
Furthermore \Rout is not fully independent of the resolution
(eg. one gets 1.41 for \nbase=500). 
The gain on \Rin \wrt the simple equiangular case is marginal.

\subsection{A quasi-optimal solution: \sa}
\label{sec:sa}
A smart pixelization unknown in astrophysics and
cosmology was proposed by mathematicians \citep{Lemaire2000}
which leads both to equal-area and close-to-square pixels. Although
relying on cube symmetries, it is not based on projecting a grid
face, but on direct spherical computations.

However the original construction algorithm is too complex to allow
a fast \angpix implementation.
In their paper, the authors exclude a simpler way to build the nodes
on regular latitude bins since it does not lead to
strictly equal-area pixels. Since this is not a strong constraint in
our case, we have tested this simplified scheme which now allows an
efficient \angpix computation. Anticipating results, 
we have nicknamed it \sa, the Similar Radius Sphere Pixelization.

\begin{figure}
  \centering
  \subfloat[]{
    \label{fig:sa_consa}
    \includegraphics[width=.5\linewidth]{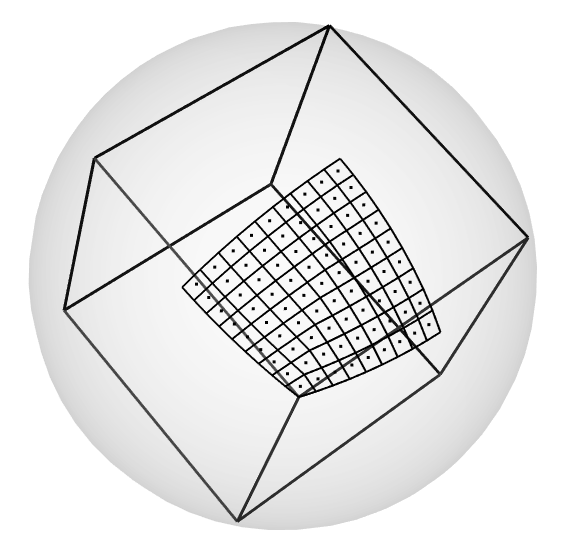}
  }
  \subfloat[]{
    \includegraphics[width=.5\linewidth]{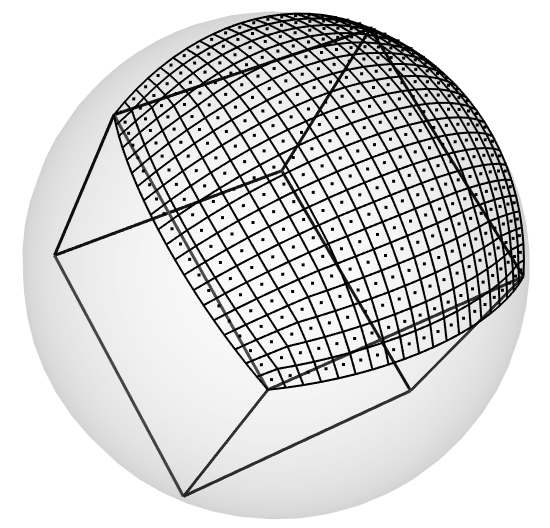}
  }
  \\
  \subfloat[]{
    \includegraphics[width=.5\linewidth]{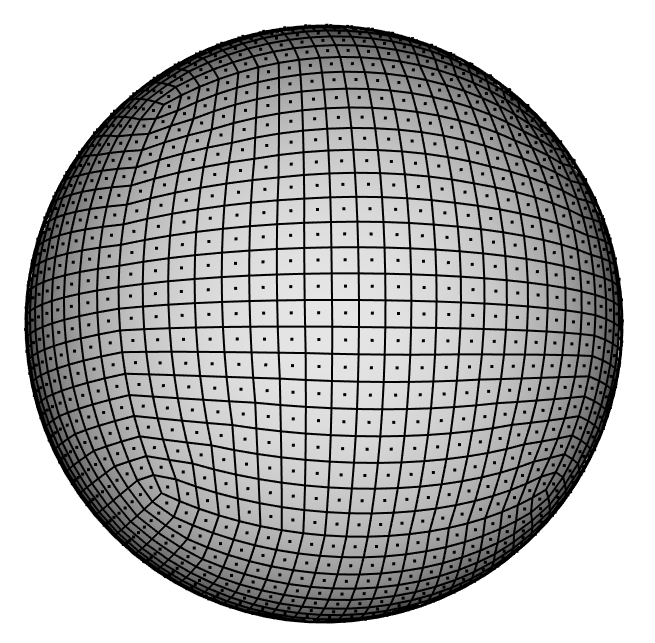}
  }
  \caption{Construction of the \sa pixelization.(a) First a quadrant
    of a projected cube face is constructed computing the node positions
    directly on the sphere.
    (b) The
    other quadrants are completed using symmetries. (c)
    The full-sphere pixelization is completed using cube rotations.}
\label{fig:sa_cons}
\end{figure}

Referring for details to the original paper, 
we present a brief summary of the main steps in the construction
process (\fig\ref{fig:sa_cons}). 
One works on a quadrant of a projected cube face and
define meridians that respect some regular triangular area spacing.
Then one regularly splits the latitudes of each meridian up to the diagonal of
the quadrant and symmetries the result \wrt it. The
construction is repeated for each quadrant and face using symmetries.

\begin{figure}
  \centering
   \includegraphics[width=\linewidth]{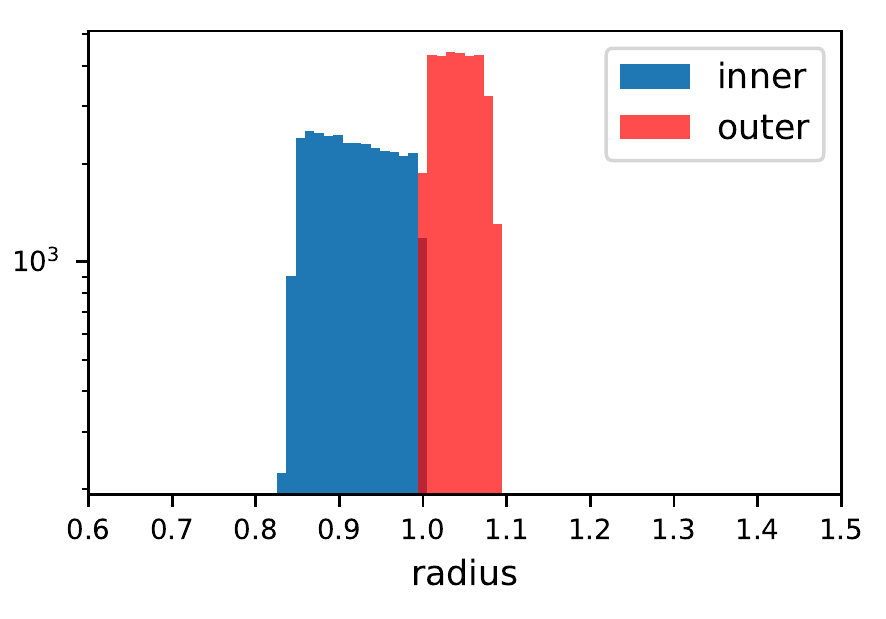}
  \caption{Histogram of the inner (in blue) and outer (in red) pixel
    radii for the \sa pixelization
    normalized to values for a square.}
    \label{fig:sa}
\end{figure}

We have constructed this pixelization and show on \fig\ref{fig:sa}
the radii distribution. We obtain
\begin{align}
\label{eq:sa}
  \dfrac{\Rin}{\Rsqin}=0.82,\quad \dfrac{\Rout}{\Rsqout}=1.10
\end{align}
There is only a 10\% departure from a square on the outer radius and
about 20\% on the inner one.

The total number of pixels is $\Npix=6~\nbase^2$ where $\nbase$,
is an \textit{even} integer\footnote{to respect symmetries used
  during the building process.} 
As for the previous cube-based pixelizations, the number of neighbors of each
pixel is 8, except for the 8 cube corners where it is 7.

\begin{table}
  \centering
  \begin{tabular}{llll}
\toprule
 parameter &    \hp &    \cs &  \sa \\
\midrule
resolution & $\nside=2^N$ & $\nbase=N$ & $\nbase=2N$ \\
\Npix & $12~\nside^2$ & $6~\nbase^2$ & $6~\nbase^2$ \\
\#neighbours & 8~(7)& 8~(7)& 8~(7) \\
\Rin & 0.65 & 0.77 & 0.82 \\
\Rout & 1.46 & 1.26 & 1.10 \\
\bottomrule
\end{tabular}
  
  \caption{ Summary of pixelization properties studied in the text.
    For the resolution parameter, $N$ represents any positive integer. The number of
    neighbors is essentially 8 in all cases; for \hp there are also
    24 pixels with 7 neighbors, while for cube-based constructions (\cs,\sa)
    there are 8 of them (at the cube corners). \Rin is the largest radius of
    the circles \textit{inscribed} on the pixels and 
    \Rout the smallest of the \textit{circumscribed} circles. They are
    normalized to
    the values for a square \ie would be exactly 1 for square pixels.
}
\label{tab:summary}
\end{table}

We summarize in Table \ref{tab:summary} the characteristics of the
pixelizations that we have studied.
According to our metric, which is to have the most homogeneous
square-like pixels with a slowly varying resolution parameter,  
\sa is clearly superior to
the others and will be our default choice both for the
\textit{joining} and \textit{reducing} indexing.

\subsection{Implementation}

We have implemented the \sa and \cs pixelizations using the \sca language
and make them publicly available in the \sparkcorr package
\footnote{The description of classes is available in the
package \texttt{docs/} directory and online at 
\safeurl{https://astrolabsoftware.github.io/SparkCorr}}
The classes provide efficient access to the main functions needed by our algorithm:

\begin{enumerate}[noitemsep]
\item \angpix :  $(\theta,\phi)\to \mathrm{index}$
\item \pixang : $\mathrm{index} \to (\theta,\phi)$
\item \neib : $\textrm{index}\to\{\textrm{indices}\}$
\end{enumerate}

We evaluate the performances of the implementations by comparing them
to \hp (\texttt{java} implementation) which is a very optimized
code. 
We shoot $10^9$ random points over the sphere
compute their index (\angpix), then the corresponding pixel centers
(\pixang) and finally the list of all the neighbors.
We have connected  the \sca classes to the \spark environment and ran on 5 workers with an
equivalent resolution parameter
of \nside=2048 for \hp and \nbase\ = 2896 for \cs and \sa. We
checked that the performances are independent of the resolution.
Table \ref{tab:perf} shows the times measured for each operation.
The \cs and \sa performances are similar to the \hp ones but for
\pixang where the latter can determine immediately the position of the
centers owing to its hierarchical decomposition but is more complicated 
for cube-based constructions.
Storing the pixel centers in a lookup table to accelerate this function
is not appropriate with \spark, since one would need to send
this table to all the executors which can be long if the table is
large due to the finite network bandwidth and can even saturate the heap memory.
Then our \pixang pure-function implementation relies on computing the
4 nodes and their barycenter for each point which in principle could
be as long as 4 times \angpix. 
The measured performances, close to the \angpix ones, 
are then quite reasonable.
The time increase of \sa over \cs simply comes from the higher
complexity of the former scheme.
In all cases, and as will be clear in the final results, these
performances are largely sufficient to not impact the full pair counting algorithm
(\sect{sec:overview}). 

\begin{table}
  \centering
  \begin{tabular}{llll}
\toprule
 operation &    \hp &    \cs &  \sa \\
\midrule
\angpix & 26s & 20s & 33s \\
\pixang & 3s & 30s & 43s \\
\neib & 10s & 15s & 15s \\
\bottomrule
\end{tabular}
  
  \caption{ Time measured (in seconds) to perform the main pixelization
    operations on $10^9$ data on 5 workers (each running 32 cores).}
\label{tab:perf}
\end{table}

Finally, we point out that using \sa we gain substantially on pair combinatorics.
As an example let us consider the data reduction for a bin width of
3.2\arcmin. With \hp the number of pixels \NpixD would be 200M (\nside=4096) while
with \sa we can get down to 34M (\nbase=2386), meaning that working
on a 100M dataset one cannot compress the data with the former but can
obtain a factor 3 of compression with the latter.
The joining pixelization is also essential for
obtaining good performances.
For instance, with a \tu=10\arcmin upper bound, the number of pixels \NpixJ, 
that decimates the pair combinatorics (\refeq{npairsX}) is around $200.10^3$ with \hp
while it is four times larger with the \sa indexing.

\section{The DES binning scenario}
\label{sec:results}

\subsection{Defining the setup}
\label{sec:setup}

We use in the following the \q{DES binning} \citep{DES:2018} which
consists of 20 bins uniformly spaced on a logarithmic scale from 2.5 to
250 arcmins and is detailed in Table \ref{tab:bin_setup}.
Once the binning is defined, we first work out the characteristics of
the joining and reducing pixelizations for each bin.

For the joining pixelization (\sect{sec:exact}) we compute the smallest possible
value of \nbase which ensures that any pair of points is at an angular distance lower
than $\Rin=\tu/2$ (see \fig\ref{fig:grille}). For \sa, using Eqs.
(\ref{eq:sa}),(\ref{eq:Rsq}) and (\ref{eq:Npix})
\begin{align}
\label{eq:npixj}
  \nbaseJ=\floor*{\sqrt{\dfrac{\pi}{6}} \dfrac{0.82}{\Rin}}
\end{align}
$\floor{}$ indicating here the $floor$ operator, possibly adding 1 if the
value is odd.

For data reduction (\sect{sec:reduce}) the pixelization only depends on the bin width ($w$).
For each bin it corresponds to the largest \nbase
value which ensures that all pixel radii are smaller than $\Rout=w/2$
(see \fig\ref{fig:boule}).
For \sa it yields
\begin{align}
\label{eq:npixd}
  \nbaseD=\ceil*{ \sqrt{\dfrac{\pi}{3}} \dfrac{1.10}{\Rout}},
\end{align}
$\ceil{}$ indicating the $ceil$ operator, possibly subtracting 1 if
the value is odd.

This defines the optimal pixelizations on each bin and thus
the total number of pixels, assuming a full sphere coverage, $\NpixD=6~\nbaseD^2$ ,
$\NpixJ=6~\nbaseJ^2$ that are indicated in Table \ref{tab:bin_setup}.

\begin{table}
  \centering
  \begin{tabular}{rrrrrr}
\toprule
 id &    \td &    \tu &  \NpixJ($10^3$) &    $w$ &  \NpixD($10^6$) \\
\midrule
  0 &   2.5 &   3.1 &  10,077.7 &  0.6 &     858.0 \\
  1 &   3.1 &   4.0 &   6,340.7 &  0.8 &     541.3 \\
  2 &   4.0 &   5.0 &   3,995.1 &  1.0 &     341.5 \\
  3 &   5.0 &   6.3 &   2,519.4 &  1.3 &     215.6 \\
  4 &   6.3 &   7.9 &   1,597.5 &  1.6 &     135.9 \\
  5 &   7.9 &  10.0 &     998.8 &  2.0 &      85.8 \\
  6 &  10.0 &  12.5 &     629.9 &  2.6 &      54.1 \\
  7 &  12.5 &  15.8 &     399.4 &  3.2 &      34.2 \\
  8 &  15.8 &  19.9 &     249.7 &  4.1 &      21.6 \\
  9 &  19.9 &  25.0 &     157.5 &  5.1 &      13.6 \\
 10 &  25.0 &  31.5 &      98.3 &  6.5 &       8.6 \\
 11 &  31.5 &  39.6 &      62.4 &  8.1 &       5.4 \\
 12 &  39.6 &  49.9 &      38.4 & 10.3 &       3.4 \\
 13 &  49.9 &  62.8 &      24.6 & 12.9 &       2.2 \\
 14 &  62.8 &  79.1 &      15.0 & 16.3 &       1.4 \\
 15 &  79.1 &  99.5 &       9.6 & 20.5 &       0.9 \\
 16 &  99.5 & 125.3 &       6.1 & 25.8 &       0.5 \\
 17 & 125.3 & 157.7 &       3.5 & 32.4 &       0.3 \\
 18 & 157.7 & 198.6 &       2.4 & 40.8 &       0.2 \\
 19 & 198.6 & 250.0 &       1.5 & 51.4 &       0.1 \\
\bottomrule
\end{tabular}
  
  \caption{Setup used with the DES binning. The first column is an
    identifier used to define bin ranges. Then comes the
    bin $[\td,\tu]$ boundaries (in arcmins) and \NpixJ the associated number of
    pixels (in thousands) of
    the joining pixelization that depends only on \tu.
    Then comes $w$ the bin width (arcmins) and its associated number of
    pixels for the data reduction \NpixD (in millions).}
\label{tab:bin_setup}
\end{table}

Note that pixel reduction is not always efficient on the first bins.
For instance, for 100M input data, there are actually more pixels in
the first 5 bins so it makes no sense using it.
One could then trigger it only when $\NpixD<\Ndata$. 

However \spark allows much more data to be processed at once. 
We work in the following on \textit{bin ranges} defined by a pair of identifiers from
Table \ref{tab:bin_setup} as $[i_{min},i_{max}]$.
We can then use the joining pixelization corresponding to
$\tu(i_{max})$ for all the range since by construction all lower
index values have a lower angular separation. 
Conversely, for data reduction (if any) we can use the pixelization
corresponding to $w(i_{min})$ since the width increases with the bin number.
Then we use some fixed-size pixelizations with $\NpixJ(_{max})$ and $\NpixD(i_{min})$ pixels,
the latter being optional.
It is important to notice that the fact that \tu and $w$ evolve one
against the other, allows treating the full histogram with very
little ranges since when \NpixJ increases \NpixD decreases and the number of
pairs becomes easily tractable (\refeq{npairsR}). In practice it 
means we need to run very little jobs on a cluster.
The optimal slicing depends on the binning, the data size and cluster
performances, but we will see that to complete the full DES histogram
on $10^8$ to $10^9$ data points we only need 2 to 4 bin ranges (\ie jobs).

\subsection{Data and infrastructure}
\label{sec:colore}
To study performances on a large and realistic dataset we used the
\colore software 
\footnote{\safeurl{https://github.com/damonge/CoLoRe}} 
that is a fast simulation populating point-like galaxies according to a
log-normal over-density distribution.
We built a catalog of around 6 billions of galaxies 
corresponding to 10 years of \LSST observations and extract
three datasets by filtering galaxies with a 
redshift between 1 and an upper cut chosen to create
different size catalogs, the \q{100M}, \q{500M}
and \q{1G} ones containing respectively $10^8, 5.10^8$ and $10^9$
point-like galaxies distributed over the full sky.

The tests were run at the \texttt{NERSC} computing center
\footnote{\safeurl{https://www.nersc.gov}} on the \spark \texttt{2.4.4} version. Each job provides to
\spark a number of sandboxed workers corresponding to the number of
nodes specified by the user 
 \footnote{\safeurl{https://docs.nersc.gov/analytics/spark}}. 
Each node has two
sockets, each socket being populated with a 2.3 GHz 16-core Haswell
processor (Intel Xeon Processor E5-2698 v3). Each node has 100 GB
memory, about 60\% being used for the cache.
Even if \texttt{NERSC} proposes high-quality supercomputer services,
we emphasize that this is not essential here since \spark is by design
a framework to be run on any data center. Some results on \spark
performances at NERSC are reported in \citet{SparkFITS:2018}.

\subsection{Performances}

\subsubsection{The \exact method}
\label{sec:unred}

\begin{figure}
  \centering
  \includegraphics[width=\linewidth]{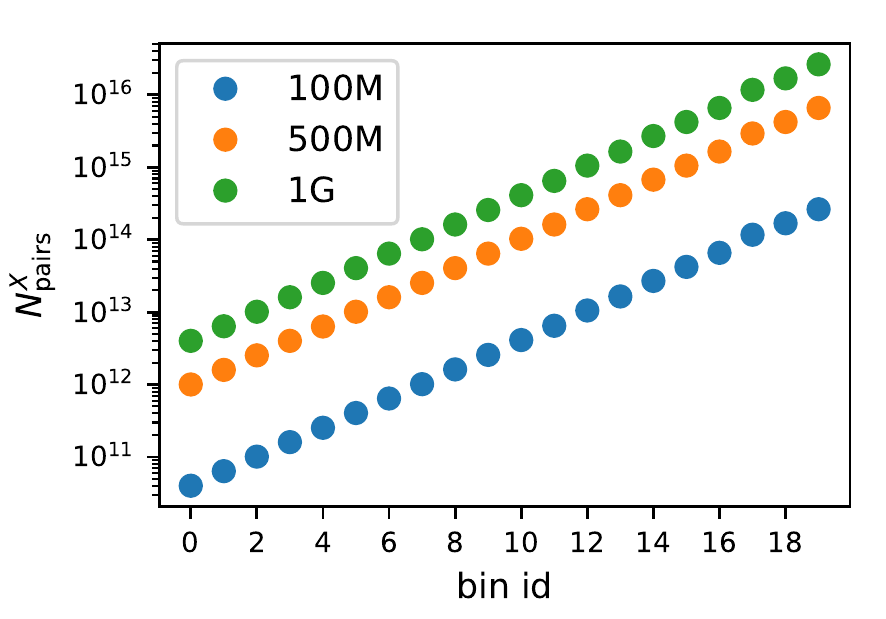}
  \caption{Theoretical number of pairs with the \exact method from bin 0
    to the shown identifier  (from Table \ref{tab:bin_setup})
    on our three datasets.}
\label{fig:npairs}
\end{figure}

The most demanding part of the algorithm is when the data reduction
cannot be applied which occurs for the first angular
distance bins, \ie when \NpixD exceeds the input data size.
The first bin range $[0,i_{max}]$ could then go up to where 
the compression starts becoming effective ($\NpixD<\Ndata$). It would lead
for example to $i_{max}=4$ on the 100M sample (Table \ref{tab:bin_setup}).
As discussed in the previous part, we can be more ambitious with
\spark and push $i_{max}$ higher.
To estimate up to where we can go, 
we consider the theoretical number of pairs
\refeq{npairsX}. Using \NpixJ from Table \ref{tab:bin_setup} 
we represent \npairsX according to $i_{max}$
on \fig\ref{fig:npairs} for the three datasets.

In order to obtain results within minutes, we found that \spark at
\texttt{NERSC} can handle up to about $10^{13}$ pairs.
By consequences, we have chosen $i_{max}=10,5$ and $1$ for the "100M", "500M"
and "1G" datasets, respectively.
Note that from this figure one can adjust the bin range to other cluster performances.

\begin{table}
  \centering
  \begin{tabular}{lcccc}
\toprule
 \Ndata &  bin range  &  nodes  &  \npairsX & wall time (mins)\\
\midrule
$10^8$ & [0,10] & 16 & $4.10^{12}$ & $1.7\pm0.1$\\
$5.10^8$ & [0,5] & 32 & $10^{13}$ & $2.4\pm0.1$\\
$10^9$ & [0,1] & 64 & $6.10^{12}$ & $1.7\pm0.1$\\
\bottomrule
\end{tabular}
  
  \caption{Timing results on the first bin range (Table \ref{tab:bin_setup}) for the three datasets  ("100M", "500M" and "1G")
    with the \exact method. We also give the number of nodes used and
    the theoretical number of pairs computed from \refeq{npairsX}.
 }
\label{tab:tomoX}
\end{table}

Table \ref{tab:tomoX} shows the timing results obtained 
running five jobs each time to measure the average and standard deviation.
Starting with 16 nodes, we doubled each time their number in
particular to hold the increasing data in the cache, which will be
discussed in more details later.
In each case a wall time of about 2 minutes is achieved.

\begin{figure}
  \centering
  \includegraphics[width=\linewidth]{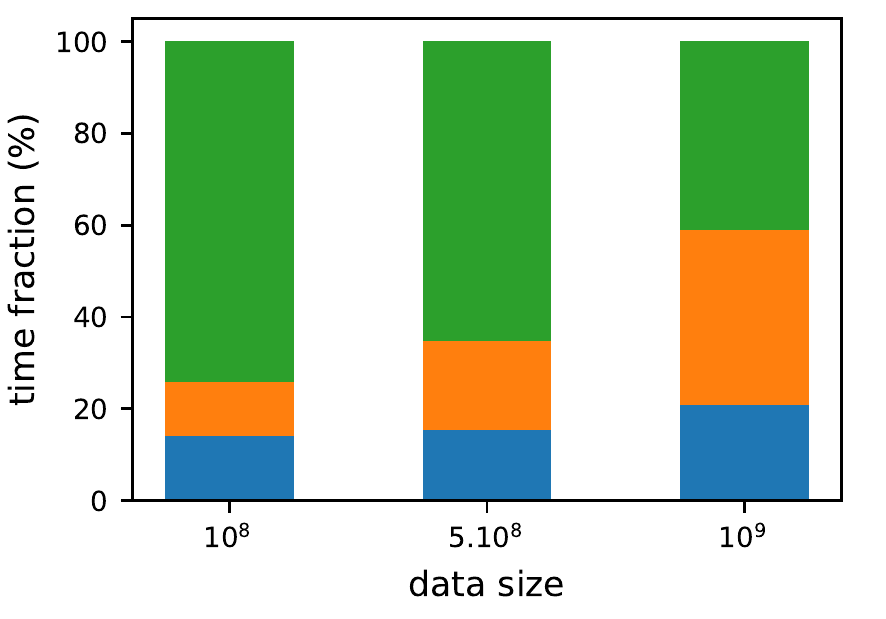}
  \caption{Relative time spent in the main steps described in the
    text of the \exact algorithm
    for Table \ref{tab:tomoX} setup.  Following \sect{sec:overview}
    terminology, in blue the \textit{source} part, in orange
  the \textit{duplicates} one and in green the \textit{join} one.}
  \label{fig:reltime}
\end{figure}

Let us investigate where time is spent. Using \sect{sec:overview} terminology,
we show on \fig\ref{fig:reltime} the relative time spend in the main parts of the algorithm.
For the "100M" and "500M" datasets most of the time is
spent in performing the combinatorics (the \texttt{join} part, point 3
in \sect{sec:overview})). For the
"1G" dataset, the \textit{duplicates} part (point 2
in \sect{sec:overview}) is increasing since
one has to repartition and put in cache about 500 GB of data and is
therefore limited by the network bandwidth.

Putting data in cache boosts the \texttt{join} part of the algorithm but one 
does not always have access to a cluster with such a high level of
in-cache memory.
So we removed the cache usage in both the \textit{source} and
\textit{duplicates} steps and measure $2.6$ mins on the "1G" dataset.
This is a minor increase \wrt $1.7$ mins obtained using the cache, 
showing that its usage is not mandatory.

\begin{figure}
  \centering
  \includegraphics[width=\linewidth]{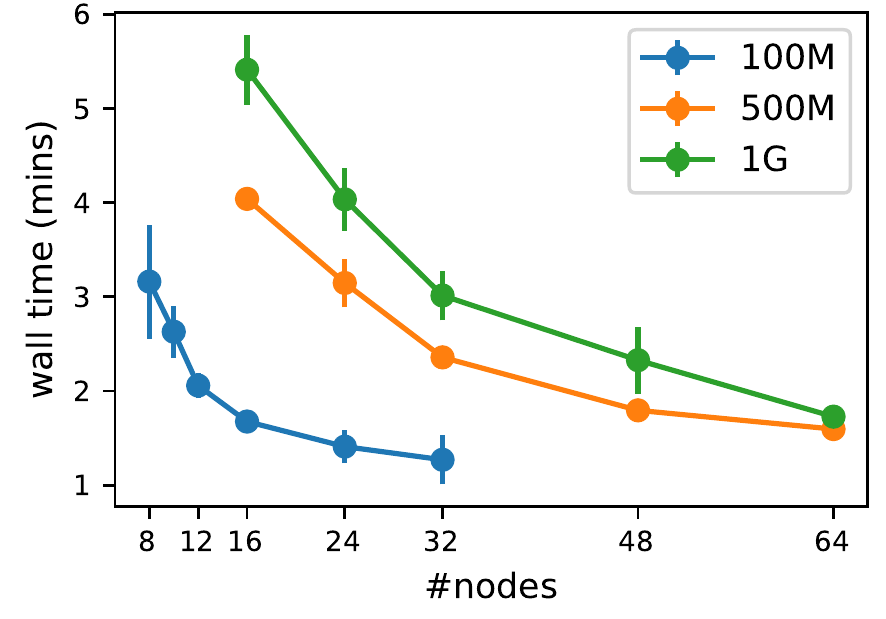}
  \caption{ Time measured on our datasets with the \exact
    method varying the number of nodes. Results are
    averaged over 5 runs for each point.}
  \label{fig:scaling}
\end{figure}

We then study how performances vary with the number of
workers (nodes). Results are presented on \fig\ref{fig:scaling}. 
The code scales nicely up to about 12 nodes on the "100M" sample and 32
for the "500M" and "1G" ones. The time is already at the 1-2 minutes
level that adding more nodes reveal some \spark overheads
(as putting data in the cache) although one still gets a small improvement.

\subsubsection{The \red method}
\label{sec:red}

When the bin number increases, \NpixJ decreases (see Table \ref{tab:bin_setup}) and combinatorics
increases too, but at the same time one can apply the data reduction
method (step 0 in sect{sec:overview})
and replace the dataset by \NpixD cells keeping the
number counts (\sect{sec:reduce}). 

Then, as in the previous section, one needs to decide the size of the bin
range that will be processed. Let us illustrate it on the "1G" dataset.
Bins 0 to 1 were already treated by the \exact method, so we start at \imin=2.
This fixes the maximal bin width and therefore the size of the
reducing pixelization. From Table \ref{tab:bin_setup} one reads \NpixD,
which is here about 340 million.
We then look for the lowest possible bin for which \NpixJ gives a number of
combinations \refeq{npairsR} that stays "small" (below $10^{13}$),
which lead us to choose \imax=5. We then proceed in the same way for
next bins. 
 
In this way, we obtain 3 bin ranges (jobs) for the "1G" dataset ([2,4],
[5,10] and [11,19]) , 2 for the "500M" one ([6,10] and [11,19]) and a single range
([11-19]) for the "100M" case.  In most cases the combinatorics is actually 
small $(\NpixD)^2/\NpixJ \ll 10^{11}$ so we use 16 nodes
everywhere but for the "1G" case for which the [2,4] index range is
slightly heavier and requires 32 nodes.

Timing results are shown in Table \ref{tab:tomoR}. With 1 to 2
minutes, they are even better
than those obtained with the \exact method

After optimization, the time spend in the \textit{Reduction} stage
(step 0 in \sect{sec:overview}) becomes negligible so that the total time  
on a given bin range becomes \textit{independent of the initial data
  size}. This can be seen in Table \ref{tab:tomoR} on the [11,19] bin
range case for which the processing time is the same whatever the
dataset size is.

\begin{table}
 \centering
 \begin{tabular}{llll}
\toprule
\Ndata  & bin range  & \npairsR &time (mins)\\
\midrule
$10^8$   & [11,19] & $8.10^{11}$ & $0.9\pm0.1$\\
\midrule
\multirow{2}{*}
{$5.10^8$} & [6,10]  & $10^{12}$ & $1.1\pm0.2$ \\ 
         & [11,19] & $8.10^{11}$ & $0.9\pm0.2$\\
\midrule
\multirow{3}{*}
{$10^9$}   & [2,4] & $3.10^{12}$ & $2.1\pm0.3$ \\
         & [5,10]  & $3.10^{12}$  & $2.3\pm0.4$ \\
         & [11,19] & $8.10^{11}$ & $0.9\pm0.1$\\
\bottomrule
\end{tabular}
  
 \caption{Timing results obtained on the three datasets with the \red
   method. Independent jobs are run over different bin ranges to finalize
   results obtained with the \exact algorithm (Table \ref{tab:tomoX}). We give the estimated
   number of pairs from \refeq{npairsR} computed from Table
   \ref{tab:bin_setup} taking for 
   \NpixD the low bin value and for \NpixJ the high one.
   Sixteen nodes were used everywhere but for the $10^9$ [2,4] case where
   it is 32.
   }
\label{tab:tomoR}
\end{table}

\subsubsection{Combination and cross-checks}
\label{sec:crosschecks}

Using both the \exact and \red independent methods (Secs.
\ref{sec:unred}, \ref{sec:red}) , one can then reconstruct
the full histogram in about 2 minutes by launching
2, 3, 4 jobs on the "100M", "500M" and the "1G" datasets, respectively\footnote{which assumes jobs run at the same time
 which is very reasonable since they are short and require little
 nodes.}. 

We undertake some tests to check the coherence of the results in
particular on the pixelization resolution side to verify that we collected all the pairs.
To do so, we use the "100M" sample and first perform a reference run with the \exact
method on a very large [0,15] bin range. The $\imax=15$ bound
being very high, the pixel size over which the
join is performed is very large (\nbaseJ\ = 40). This run took about 8 minutes on
64 nodes.
Then, we compare the number of pairs reconstructed up to bin 10
against our default implementation that fixes the pixelization
according to \refeq{npixj} which had a much smaller resolution (\nbaseJ=128)
Results are shown in Table \ref{tab:xvsx} and indicate a very good
agreement between the two methods: they are exactly the
same up to bin 8 and differ by $10^{-6}$ for bin 10, 
showing that our joining pixel size (\refeq{npixj}) is well adapted to the problem.

\begin{table}
 \centering
 \begin{tabular}{rrr}
\toprule
 id &      $N_{ref}$ &    $N_{def}^\exact$ \\
\midrule
  0 &    506727642 &    506727642 \\
  1 &    800041282 &    800041282 \\
  2 &   1260509592 &   1260509592 \\
  3 &   1979959475 &   1979959475 \\
  4 &   3096818992 &   3096818992 \\
  5 &   4815924052 &   4815924052 \\
  6 &   7438665544 &   7438665544 \\
  7 &  11411877241 &  11411877241 \\
  8 &  17425875136 &  17425875136 \\
  9 &  26591445306 &  26591444936 \\
 10 &  40723594728 &  40723260295 \\
\bottomrule
\end{tabular}
  
 \caption{Number of pairs measured on the $10^8$ sample with the
   \exact method with the \textit{reference} and \textit{default} setup}
\label{tab:xvsx}
\end{table}

We know that the compression applied in the \red method is not
absolutely lossless (\refeq{binning}). There is a slight smearing around
the binning boundaries which is difficult to predict.
In order to study this effect, two jobs have been launched 
with the \red method on the [5,10] and [10,15] bin ranges. 
We then compare the values obtained with those of our reference run
that is \exact. The results shown in Table \ref{tab:xvsr} shows
that the precision of the \red method is at the $10^{-3}$ level

\begin{table}
 \centering
 \begin{tabular}{lllr}
\toprule
 id &       $N_{ref}$ &      $N_{def}^\red$ &  rel. error \\
\midrule
  5 &    4815924052 &    4820614147 & -0.00097 \\
  6 &    7438665544 &    7436640102 &  0.00027 \\
  7 &   11411877241 &   11353806206 &  0.00509 \\
  8 &   17425875136 &   17448008463 & -0.00127 \\
  9 &   26591445306 &   26579894562 &  0.00043 \\
 10 &   40723594728 &   40730263699 & -0.00016 \\
\midrule
 10 &   40723594728 &   40767727583 & -0.00108 \\
 11 &   62746460926 &   62752402410 & -0.00009 \\
 12 &   97293945900 &   96819343446 &  0.00488 \\
 13 &  151684971831 &  151890299528 & -0.00135 \\
 14 &  237573071187 &  237474741141 &  0.00041 \\
 15 &  373467108791 &  373494164330 & -0.00007 \\
\bottomrule
\end{tabular}
  
 \caption{Number of pairs measured with the reference setup and two
   \red runs ([5,10] and [10,15] bin ranges in Table
   \ref{tab:bin_setup}) and relative fraction between them.
}
\label{tab:xvsr}
\end{table}

\section{Conclusion}

We have shown how to perform pair-counting on a standard
logarithmically binned histogram, up to one billion input data in about two
minutes using the \spark big data technology. The algorithm we developed
uses standard functions on dataframes and was shown to scale with the number of nodes. 
This demonstrate that \spark, although not very used in the
astrophysics community, is a valuable tool to handle very large combinatorics.

To achieve this level of performance we have revisited the 
\textit{non}-hierarchical sphere pixelization
and propose a new one, \sa(\sect{sec:sa}) that indexes the sky with
regular shape quadrangles. 
While \hp is widely used in astrophysics, \sa is much more adapted
to all operations involving some local search over the sphere.

Although some sate-of-the-art HPC implementation \JP{of pair counting algorithm} (as \JP{is done in} \texttt{treecorr}) 
can achieve similar performances, they do it thanks to an evolved
\textit{algorithm} (and approximations) not \textit{data structure}. There are two advantages in exploring
a \spark implementation. 
First, the big data approach allows an optimized `` brute-force'' 
computation up to quite large separation angles, meaning the results
are \textit{exact}. Once the full set of pairs is reconstructed  the binning
process itself is light, meaning that any binning (as a linear one)
gives similar performances \footnote{which is not the case when using
 a recursive balltree algorithm. Linear binings for instance are very limited.}
Second, the main difference with a state-of-the-art \JP{implementation} written
by expert(s) is that the data structure (a dataframe) and algorithm
(indexing on the sphere and join) is much simpler.
With little effort and no knowledge about parallel computing, 
any user can obtain performances similar to the most specialized
software on this high combinatorics problem.
With little adaptation one can also address other problems related to
local distance computations as:
\begin{itemize}
\item k-NN nearest neighbor search, 
\item cross-match between two catalogs,
\item cluster building with some linking length (as the `` friend of
  friends'' method used for instance in N-body simulations to reconstruct halos). 
\end{itemize}
And of course the method can be adapted straightforwardly to the
Euclidian case where indexing is provided simply by binning each dimension.
Although we presented \JP{timing} results on a super-computer we
remind that it is not a necessary requirement since \spark is in essence designed
to be run on any set of connected computers.

\section*{Acknowledgements}
SP thanks Bertrand Maury for indicating us the \citet{Lemaire2000} paper.
We acknowledge the use of the \hp package
\citep{2005ApJ...622..759G}.
The software was run at the National Energy Research Scientific
Computing Center, a DOE Office of Science User Facility supported by
the Office of Science of the U.S. Department of Energy under Contract
No. DE-AC02-05CH11231. Resources were obtained through the LSST Dark
Energy Science Collaboration, within which the authors are exploring
applications of this work.

\section*{Data availability}

The data used to produce the results as described in
Sect.\ref{sec:colore} are publically available at
\safeurl{https://me.lsst.eu/plaszczy/scalingpaircount}.
They consist in 4 tar-compressed \texttt{parquet} files:
\begin{verbatim}
tomo1M.parquet.tar, tomo100M.parquet.tar, 
tomo500M.parquet.tar, tomo1GB.parquet.tar
\end{verbatim}
corresponding respectively to the $10^6, 10^8$, $5.10^8$ and $10^9$ sample datasets. 
The complete size is of 12 GB. 
\change{
For an initiation on a personal laptop, you can just grab and untar the \texttt{tomo1M.parquet} file  and follow for instance the example of \sect{sec:sql}. You need a working \texttt{pyspark} implementation which can be downloaded from \safeurl{https://spark.apache.org/downloads.html}.
}
\appendix

\section{Pixelization properties}
\label{sec:app}
We give in this appendix more details about the cube-based pixelizations
discussed in the text.
In each case we have built the nodes of the pixelization for one face
(the other being equivalent by symmetry)  
and study some properties of the pixels determined from the four
delimiting nodes :
\begin{itemize}
\item Area,
\item ellipticity $\left|\tfrac{p-q}{p+q}\right|$ , $p,q$ being the
  diagonal lengths,
\item outer radius (of the circumscribed circle) ,
\item inner radius (of the inscribed circle).
\end{itemize}
In all cases we used exact formulas for quadrilaterals.  The
area and ellipticity are deeply related to the radii since having a
large area together with a large ellipticity directly impacts the
distance of the center to its corners, thus the outer radius.

\subsection{\cs}
\label{app:cs}
\begin{figure}
  \centering
  \subfloat[]{
    \includegraphics[width=.5\linewidth]{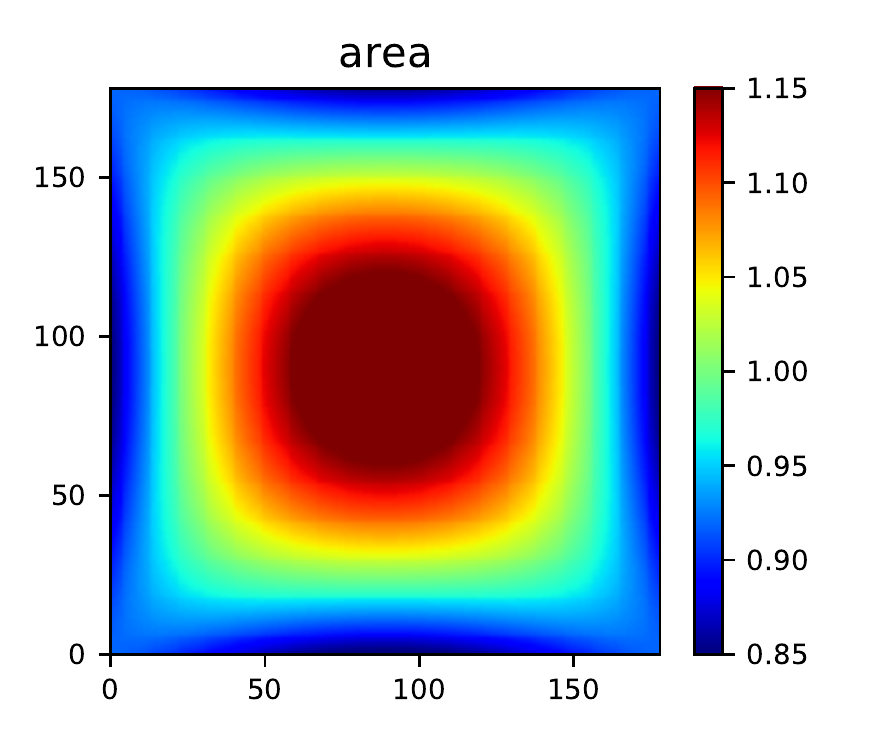}
  }
  \subfloat[]{
    \includegraphics[width=.5\linewidth]{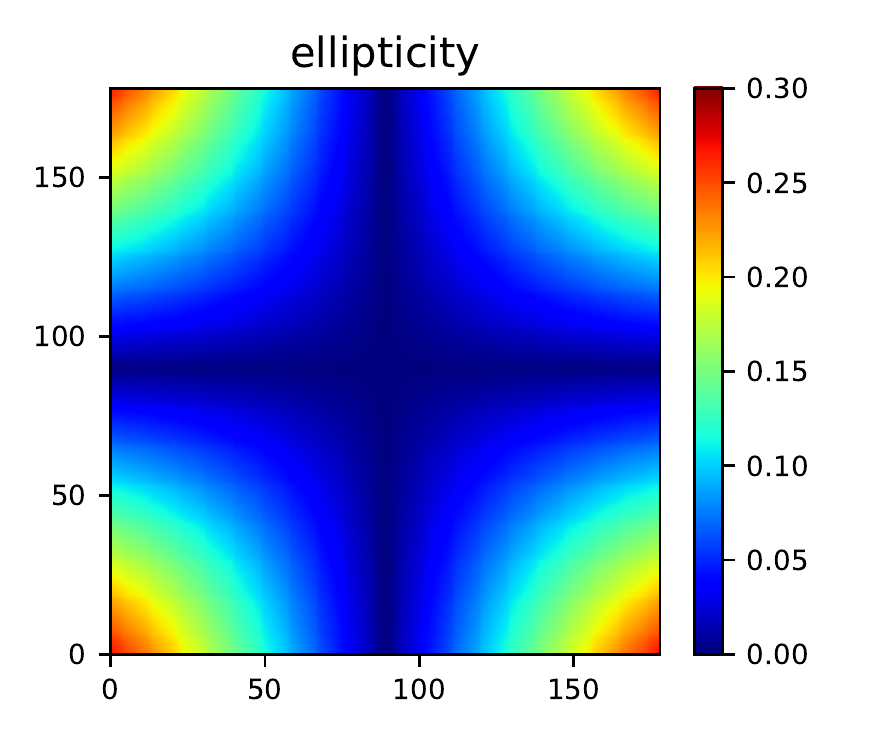}
  }\\
  \subfloat[]{
    \includegraphics[width=.5\linewidth]{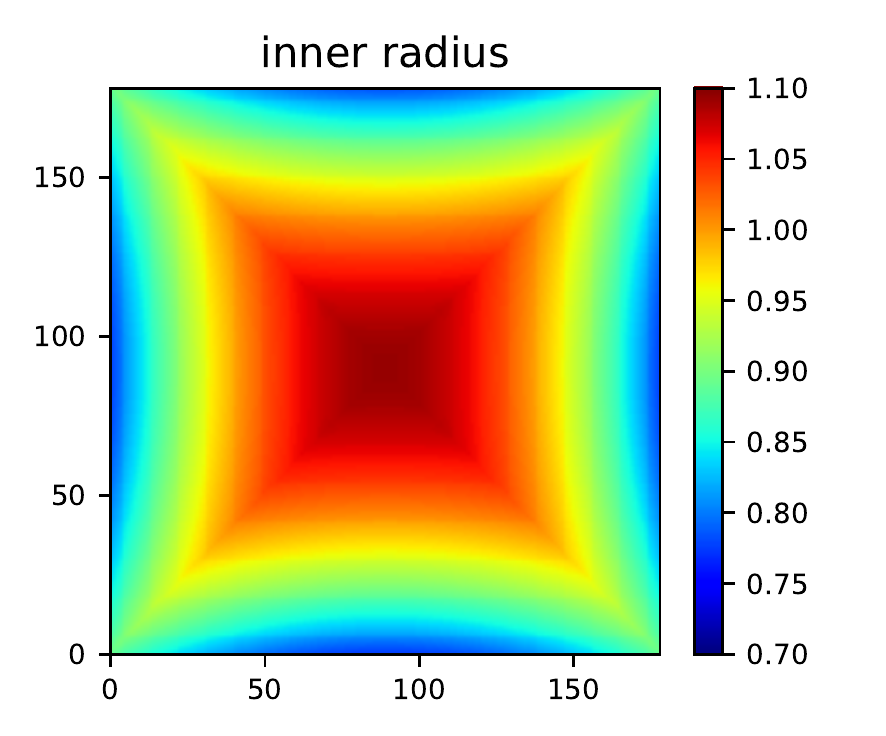}
  }
  \subfloat[]{
    \includegraphics[width=.5\linewidth]{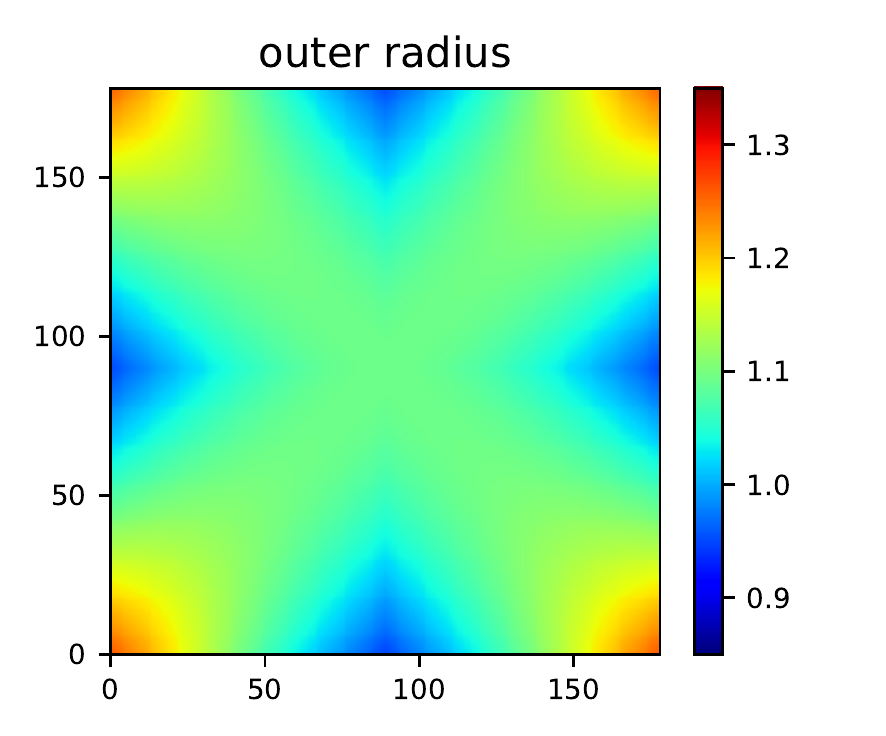}
  }
  \caption{Properties of the \cs pixels. Indices run
    over one face of the projected cube (\nbase=180) and values are normalized to
    those of a square \ie would be exactly 1 for the area, inner and outer
    radii, 0 for ellipticity.
    }
\label{fig:prop_cubed}
\end{figure}

\fig\ref{fig:prop_cubed} shows the results of the \cs pixelization
for \nbase=180. We normalize all our values to those of squares (\refeq{Rsq})
so that \Rin, \Rout and the area would be
exactly 1 for square pixels and of ellipticity 0.

The pixel area varies by 10\% and decreases near the corners. The
ellipticity increases near the corners corner.

\subsection{\cobe}
\label{app:cobe}

\begin{figure}
  \centering
  \subfloat[]{
    \includegraphics[width=.5\linewidth]{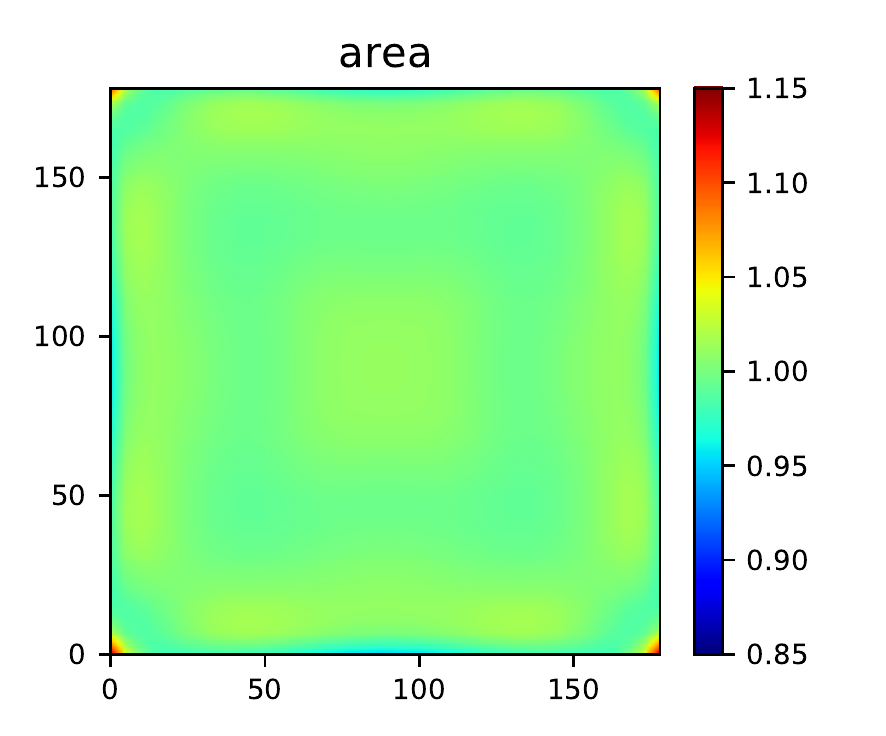}
  }
  \subfloat[]{
    \includegraphics[width=.5\linewidth]{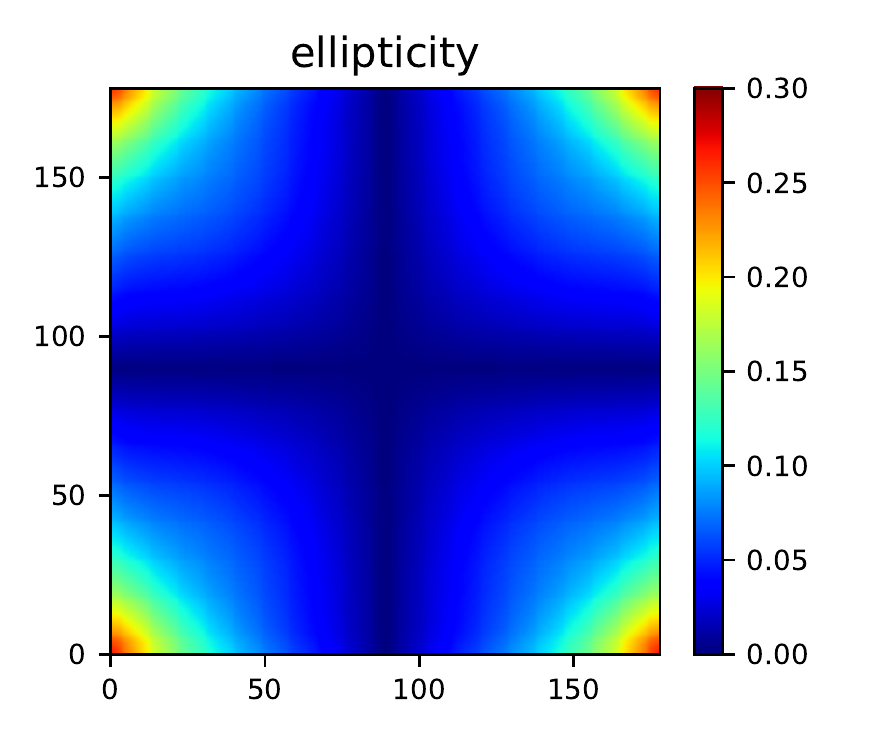}
  }\\
  \subfloat[]{
    \includegraphics[width=.5\linewidth]{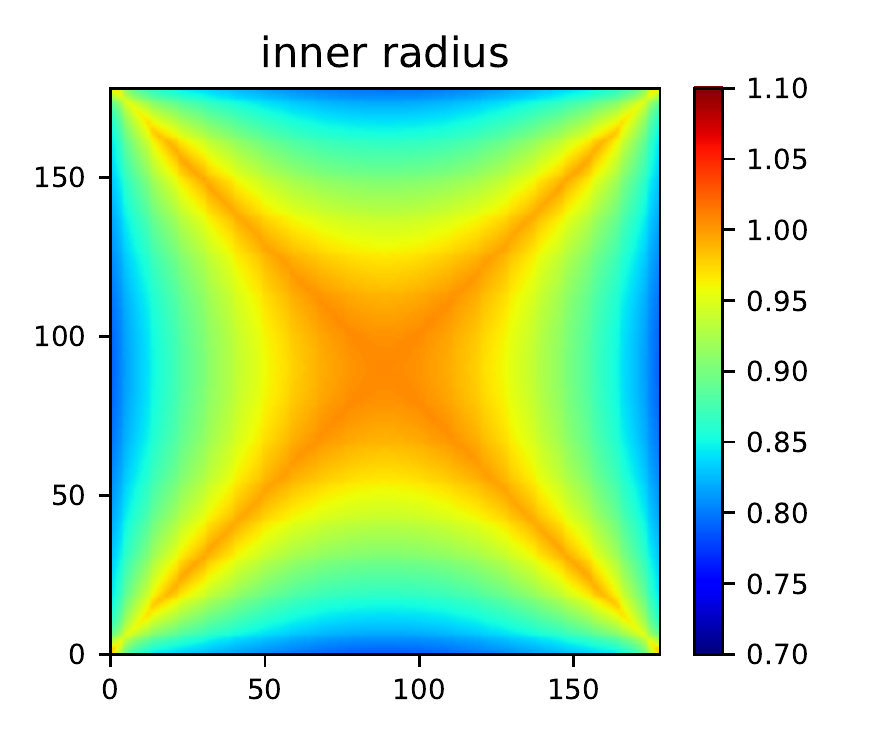}
  }
  \subfloat[]{
    \includegraphics[width=.5\linewidth]{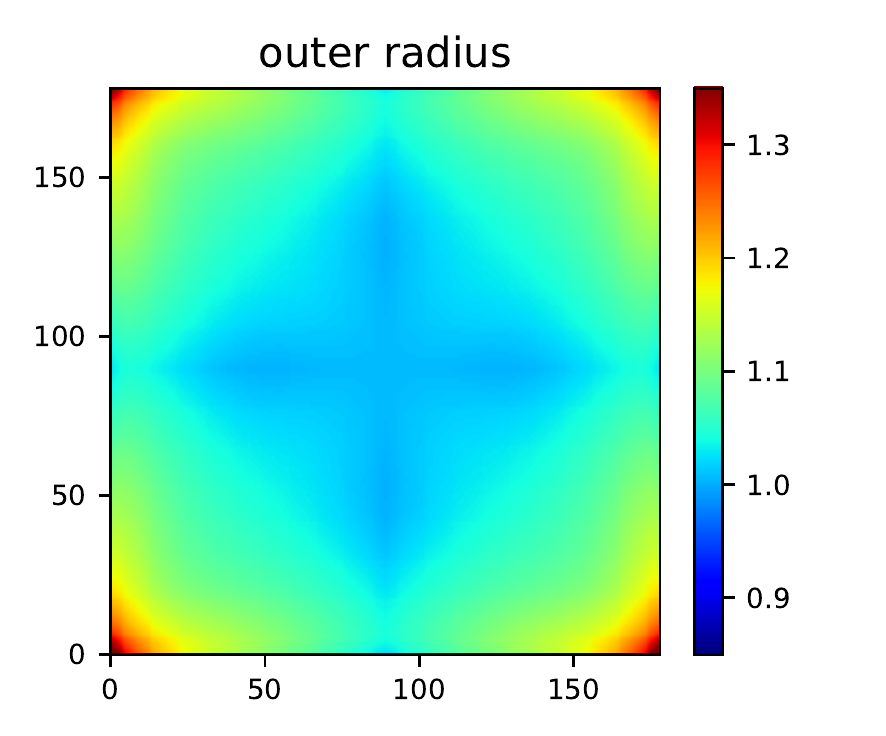}
  }
  \caption{Properties of the \cobe pixels. Indices run
    over one face of the projected cube (\nbase=180) and values are normalized to
    those of a square \ie would be exactly 1 for the area, inner and outer
    radii, 0 for ellipticity.
}
\label{fig:prop_cobe}
\end{figure}

The same properties are studied on \fig\ref{fig:prop_cobe} using the \cobe mapping of the
points on the face which was designed to obtain almost equal areas.

Indeed the pixel area becomes very uniform (below 1\% as was claimed)
\textit{but near the corners} where a $\simeq7$\% excess is observed.
The ellipticity is also globally closer to 0 but also increases near
the corners. The product of both effects lead to a dramatic increase in the
outer radius near the cube corners that was observed on \fig\ref{fig:co}.
The inner radius is globally better than on \fig\ref{fig:prop_cubed} 
but its lowest value (\Rin) is about the same than for \cs.

\subsection{\sa}
\label{app:sa}

\begin{figure}
  \centering
  \subfloat[]{
    \includegraphics[width=.5\linewidth]{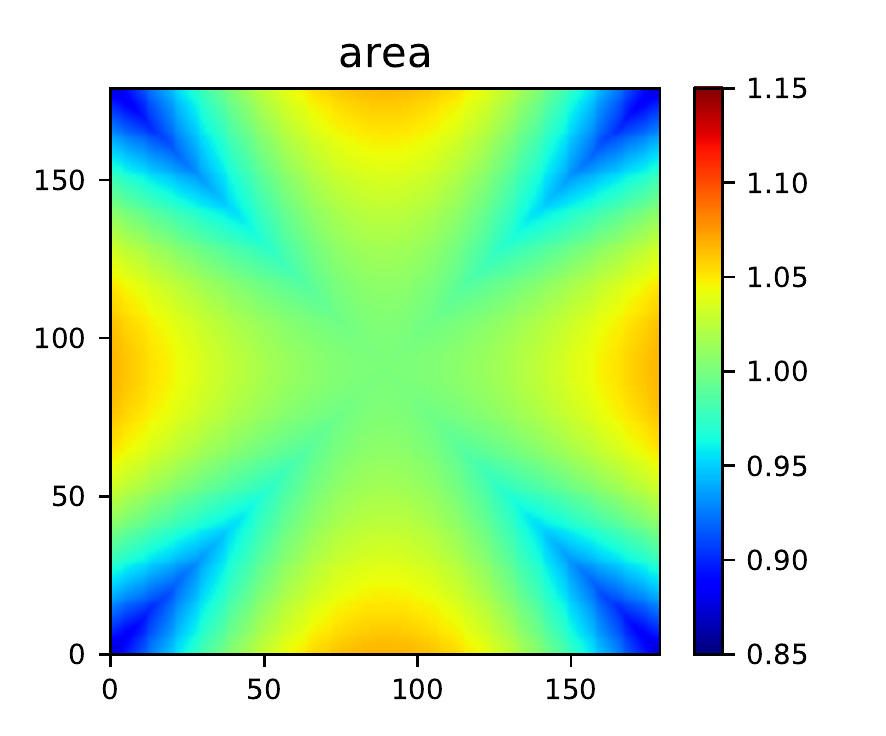}
  }
  \subfloat[]{
    \includegraphics[width=.5\linewidth]{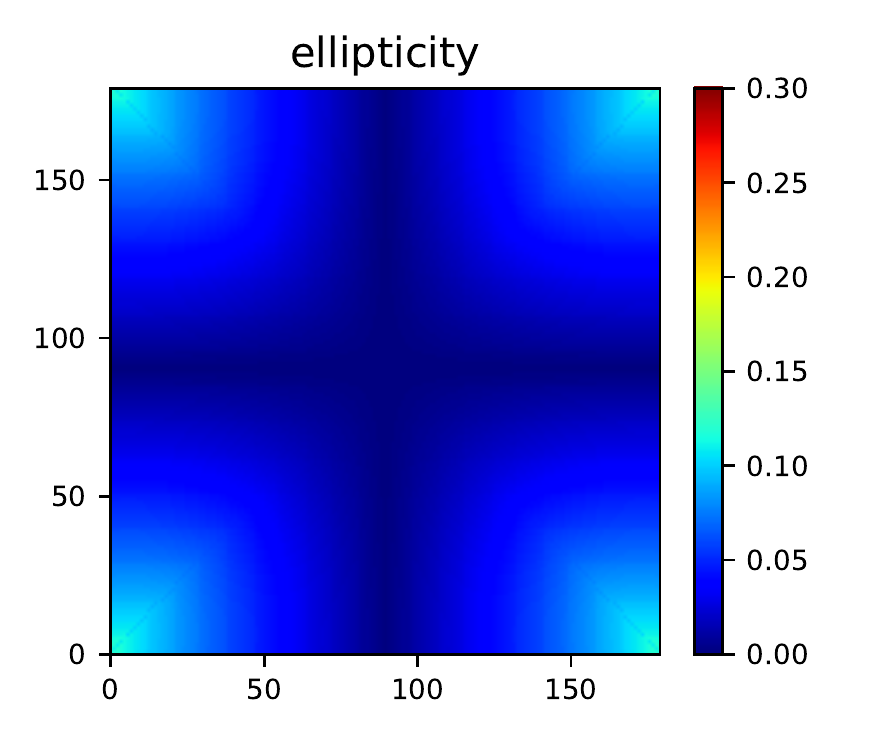}
  }\\
  \subfloat[]{
    \includegraphics[width=.5\linewidth]{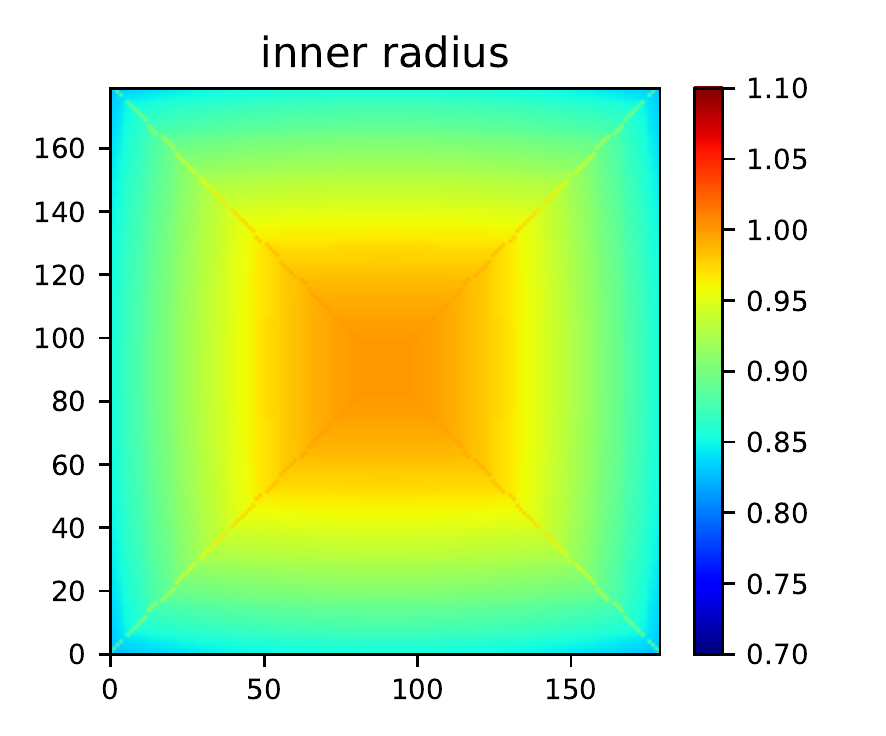}
  }
  \subfloat[]{
    \includegraphics[width=.5\linewidth]{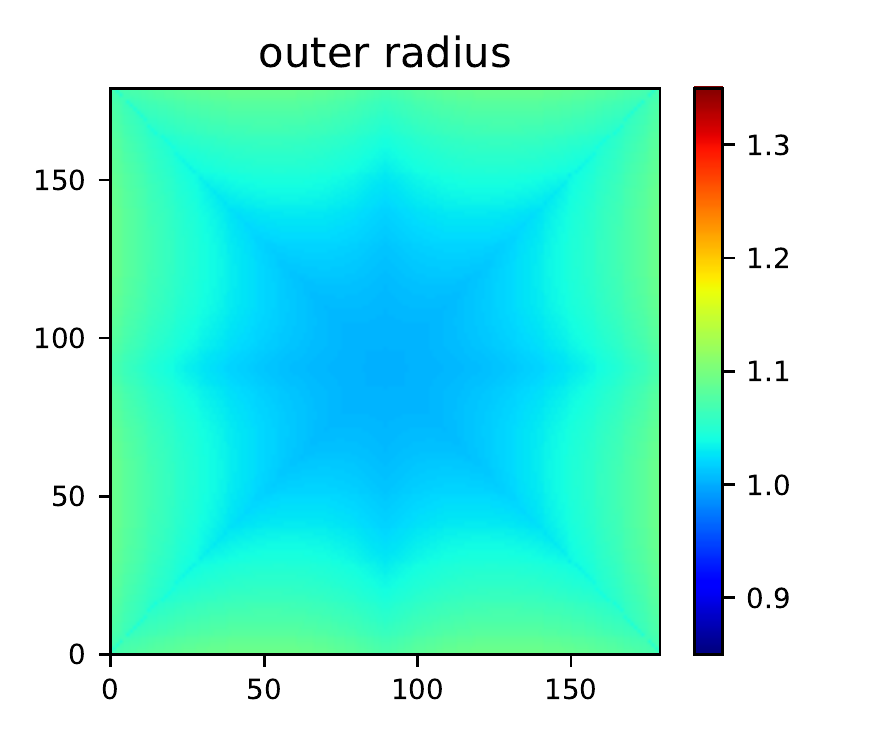}
  }
  \caption{\sa properties. Indices run
    over the face of a (\nbase=180) and values are normalized to
    those of a square \ie would be exactly 1 for the area, inner and outer
    radii, 0 for ellipticity}
\label{fig:prop_sa}
\end{figure}

\sa is a very different pixelization that does \texttt{not} project
points of the cube to the sphere but performs directly spherical computations.
However since it relies on cube symmetries, one can still index pixels
in the same way as previously.
We show the results on \fig\ref{fig:prop_sa}. 

The pixel area is less uniform than in the \cobe case but
\textit{decreases} near the corner. The ellipticity is excellent and
interestingly slightly \textit{increases} near the corner: the product
of both effects leads to an excellent outer radius that does not
increase on the corners. The inner radius is also very good, the cross
appearing here being due to the symmetrization along the diagonals 
which can be spotted on \fig\ref{fig:sa_consa}.

\newpage
\bibliographystyle{aasjournal} 
\bibliography{SparkCorr}

\end{document}